\documentclass{IEEEtran}

\usepackage{cite}
\usepackage{graphicx}
\usepackage[cmex10]{amsmath}
\usepackage{amssymb}
\usepackage{algorithmic,algorithm}
\usepackage{url}
\usepackage{subfig}

\usepackage{pgfplots}
\usepackage{tikz}
\usepackage{stmaryrd}

\usepackage{bm} 

\DeclareMathOperator*{\argmin}{arg\,min}
\DeclareMathOperator*{\argmax}{arg\,max}

\DeclareMathOperator{\acos}{acos}
\DeclareMathOperator{\R}{\mathrm{Re}}

\newcommand{\figref}[1]{Fig.~\ref{#1}}

\newcommand{\eqnref}[1]{Eqn.~(\ref{#1})}

\newcommand{\secref}[1]{Section~\ref{#1}}
\newcommand{\algoref}[1]{Algorithm~\ref{#1}}

\newcommand\va{\ensuremath{\mathbf{a}}}
\newcommand\vb{\ensuremath{\mathbf{b}}}
\newcommand\vc{\ensuremath{\mathbf{c}}}

\newcommand\vf{\ensuremath{\mathbf{f}}}
\newcommand\vg{\ensuremath{\mathbf{g}}}

\newcommand\vn{\ensuremath{\mathbf{n}}}

\newcommand\vt{\ensuremath{\mathbf{t}}}
\newcommand\vu{\ensuremath{\mathbf{u}}}
\newcommand\vv{\ensuremath{\mathbf{v}}}
\newcommand\vw{\ensuremath{\mathbf{w}}}
\newcommand\vx{\ensuremath{\mathbf{x}}}
\newcommand\vy{\ensuremath{\mathbf{y}}}

\newcommand\valpha{\ensuremath{\bm{\alpha}}}
\newcommand\vbeta{\ensuremath{\bm{\beta}}}
\newcommand\vgamma{\ensuremath{\bm{\gamma}}}

\newcommand\mA{\ensuremath{\mathbf{A}}}
\newcommand\mB{\ensuremath{\mathbf{B}}}
\newcommand\mC{\ensuremath{\mathbf{C}}}
\newcommand\mD{\ensuremath{\mathbf{D}}}
\newcommand\mE{\ensuremath{\mathbf{E}}}
\newcommand\mF{\ensuremath{\mathbf{F}}}
\newcommand\mG{\ensuremath{\mathbf{G}}}

\newcommand\mU{\ensuremath{\mathbf{U}}}
\newcommand\mV{\ensuremath{\mathbf{V}}}

\newcommand\mPsi{\ensuremath{\bm{\Psi}}}

\begin{document}
\title{Compressive Parameter Estimation\\for Sparse Translation-Invariant
Signals\\Using Polar Interpolation}
\author{Karsten~Fyhn,
        Marco~F.~Duarte,~\IEEEmembership{Senior Member,~IEEE,}
        and~S\o ren~Holdt~Jensen,~\IEEEmembership{Senior Member,~IEEE}
\thanks{Copyright (c) 2014 IEEE. Personal use of this material is permitted. However, permission to use this material for any other purposes must be obtained from the IEEE by sending a request to pubs-permissions@ieee.org.}%
\thanks{The authors would like to thank Dr. Tobias Lindstr\o m Jensen of Aalborg
    University for many helpful
    discussions about the convex optimization problem formulation. This work is
    supported by The Danish Council for Strategic Research under grant number
    09-067056, Danish Center for Scientific Computing and by a EliteForsk travel scholarship under grant number
    11-116371. }
\thanks{ K. Fyhn and S. H. Jensen are with the Department of Electronic Systems,
    Aalborg University, Denmark.  E-mail: \{kfn,shj\}@es.aau.dk.} 
\thanks{M. F. Duarte is with
    the Department of Electrical and Computer Engineering, University of Massachusetts,
    Amherst, MA 01003, USA.  E-mail: mduarte@ecs.umass.edu.}
}

\maketitle

\begin{abstract}
We propose new compressive parameter estimation algorithms that make use of
polar interpolation to improve the estimator precision. Our work extends previous 
approaches involving polar interpolation for compressive parameter estimation in two 
aspects: ($i$) we extend the formulation from real non-negative amplitude parameters to arbitrary complex ones, and ($ii$) we 
allow for mismatch between the manifold described by the parameters and its 
polar approximation. To quantify the improvements afforded by the proposed extensions, 
we evaluate six algorithms for estimation of parameters in sparse translation-invariant
signals, exemplified with the time delay estimation problem. The evaluation is
based on three performance metrics: estimator precision, sampling rate and
computational complexity. We use compressive sensing with all the algorithms to
lower the necessary sampling rate and show that it is still possible to attain
good estimation precision and keep the computational complexity low. 
Our numerical
experiments show that the proposed algorithms outperform existing approaches that 
either leverage polynomial interpolation or are based on a conversion to a 
frequency-estimation problem followed by a super-resolution algorithm. 
The algorithms studied here provide various tradeoffs between computational 
complexity, estimation precision, and necessary sampling rate. The work shows 
that compressive sensing for the class of sparse translation-invariant signals allows 
for a decrease in sampling rate and that the use of polar interpolation increases the estimation precision.
\end{abstract}


\section{Introduction}
\IEEEPARstart{C}{ompressive} sensing (CS) is a technique to simultaneously acquire and reduce the
dimensionality of sparse signals in a randomized fashion. More precisely, in the
CS framework, a signal $\vf \in \mathbb{C}^N$ is sampled by $M$ linear
measurements of the form $\vy = \mA \vf$, where $\mA \in \mathbb{C}^{M \times
N}$ is a sensing matrix and $M < N$. In practice, the measurements are acquired
in the presence of additive signal and measurement noise $\vn$ and $\vw$, respectively,
in which case we have $\vy = \mA\left(\vf + \vn\right) + \vw$.

In many applications, the signal $\vf$ is not sparse but has a sparse
representation in some dictionary $\mD\in\mathbb{C}$. In other words, we have
$\vf = \mD \vx$, where $\vx\in\mathbb{C}$ is $K$-sparse (i.e., $\|\vx\|_0 \leq
K$). Under certain conditions on the matrix $\mA$
\cite{Donoho2006,CandesRombergTao2005}, we can recover $\vx$ from the
measurements $\vy$ through the following $\ell_1$-minimization problem (which we
refer to as \textit{$\ell_1$-synthesis}):
\begin{align}
    \hat{\vx} = \arg \min_{\tilde{\vx} \in \mathbb{C}^N} \|\tilde{\vx}\|_1
    ~~\textrm{s.t.}~~ \|\mA \mD \tilde{\vx} - \vy\|_2 \leq \epsilon,
	\label{eqn:synthesis}
\end{align}
where $\epsilon$ is an upper bound on the noise level $\|\mA\vn+\vw\|_2$.  Note
that optimal recovery of $\vx$ from the optimization in \eqnref{eqn:synthesis}
is guaranteed only when the elements of the dictionary $\mD$ form an orthonormal
basis, and thus are incoherent \cite{CandesRomberg2007,Rauhut2008}. However, in
many applications, the signal of interest is sparse in an overcomplete
dictionary or a frame, rather than in a basis. 

Classic CS, as defined in \cite{Donoho2006,CandesRombergTao2005}, requires sparsity in some matrix dictionary to work, but in many
cases a signal may be sparse with respect to some parametric model instead.
Previous work has shown that CS may experience problems in such cases, when
using the traditional dictionary-based approach \cite{Nielsen2012}. One
such class of signals is sparse translation-invariant signals. Here,
translation invariance or translation symmetry refers to the Euclidian norm of
the signal, which must remain the same after translation. In this work, a translation-invariant signal is defined as follows. Let $\vg(b_i) =
\mathcal{M}_{\vg}(b_i) \in \mathbb{C}^N$ denote a point in the signal manifold
$\mathcal{M}_{\vg}(\cdot)$ parameterized by a
translation parameter $b_i$.  A function $\vg(b_i)$ is translation-invariant if
it fulfills two requirements: 1) preservation of the $\ell_2$-norm under
parameter translation, $\|\vg(b_1)\|_2 = \|\vg(b_2)\|_2, \forall ~b_1, b_2$, and 2) locally
constant (symmetrical) curvature of the signal manifold, $\|\vg(b) -
\vg(b-\Delta)\|_2 = \|\vg(b) - \vg(b+\Delta)\|_2, \forall~b$, where
$\Delta$ is some sufficiently small change in the parameter. 

Now define the signals of interest as:
\begin{align}
	\vf(\va, \vb) = \sum_{k=1}^K a_k \vg(b_k),
    \label{eqn:signal_model}
\end{align}
where $K$ is the number of signal components, $\va = \begin{bmatrix} a_1 & a_2 &
\ldots & a_K \end{bmatrix} \in \mathbb{C}^{1\times K}$ is a vector of complex amplitude coefficients,
$\vg(b)$ is a translation-invariant parametric signal, parameterized by a
translation parameter from $\vb = \begin{bmatrix} b_1 & b_2 & \ldots & b_K
\end{bmatrix} \in \mathbb{R}^{1\times K}$. The problem then becomes to estimate $\va$ and $\vb$.

Two examples of estimation problems with translation-invariant signals are Time Delay
Estimation (TDE) and Frequency Estimation (FE). TDE of one or more known signal
waveforms from sampled data is of interest in several fields such as radar,
sonar, wireless communications, audio, speech and medical signal processing. The
TDE problem is often defined as receiving a known signal with an unknown delay
and amplitude coefficient that must be estimated. Similarly, FE concerns the
estimation of the frequency components of a received sum of complex exponentials, which
is of interest in seismology, audio, speech and music processing, radar and
sonar. 

For this type of estimation problems, there can be different parameters and
performance metrics. In this work, we focus on three important performance
metrics: estimation precision, computational complexity, and necessary sampling
rate to acquire the analog signal. We use CS to lower the necessary sampling
rate while minimizing the corresponding loss in estimation precision. The algorithms we evaluate
vary in computational complexity and, not surprisingly, the most computationally
heavy algorithms feature the best performance. In some cases, the difference between 
the best and worst algorithms' estimation precision performance is four orders of
magnitude, while the computational complexity difference is two orders of magnitude larger.
It follows that this becomes a design trade-off for individual problems.

We propose two algorithms that leverage \emph{polar interpolation}, as introduced in \cite{Ekanadham2011}, to improve
the estimation precision. Interpolation is necessary because of the required
discrete dictionary in CS systems. With a parametric dictionary matrix $\mD$, we assume the delay or
frequency parameter takes values from a finite set only:
\begin{align}
    \mD = \begin{bmatrix} \vg(b_1) & \vg(b_2) & \cdots & \vg(b_N) \end{bmatrix}.
    \label{eqn:dictionary}
\end{align}
In reality, the parameter is drawn from a continuous interval.  One way to
overcome this is to increase the number of atoms in the dictionary; however,
this increases the coherence. Instead, the proposed algorithms feature a
dictionary that can sparsely represent any sparse translation-invariant signal
with a parameter drawn from a continuous interval. Polar interpolation was shown to outperform the more popular parabolic interpolation scheme in~\cite{Ekanadham2011}. This is due to the fact that it uses more prior knowledge about the problem at hand (such as translation invariance) than other interpolation approaches. We elaborate on these properties in Section~\ref{sec:polar_interpolation}.
 
In a recent paper, it was shown how polar interpolation may be utilized for
FE in the case where the amplitude coefficients are real and non-negative
\cite{Fyhn2013}. In this paper, we will show that if arbitrary 
complex amplitude coefficients are allowed (positive and negative as well as real and imaginary), then the coherence introduced by the FE
dictionary to enable interpolation does not allow for a unique, sparse solution.
We postulate that this broader problem may be solved with stronger constraints
or a different convex optimization formulation, but we do not focus on this
extension. Instead we have our main focus on the use of interpolation and CS to
solve the TDE problem where such coherence is not an issue.

The contribution of this paper consists of mainly two points: 1) We extend the work from \cite{Ekanadham2011} and propose two 
polar interpolation-based TDE algorithms that outperform other TDE algorithms, 
but may also be used for other problems involving sparse translation-invariant 
signals, and 2) We evaluate the performance of a suite of different TDE algorithms 
when coupled with CS.

In the next section, we present previous work in the area of interpolation
between dictionary elements and TDE estimation on a continuous parameter space.
In \secref{sec:polar_interpolation} we review the polar interpolation technique
and introduce an advanced convex optimization formulation to handle coefficient
vectors that are not real and non-negative. This is followed by \secref{sec:ibomp}
in which we introduce an iterative, greedy algorithm based on interpolation which may be followed by an optional convex optimization solver to improve the solution further. In
\secref{sec:numerical_experiments} we evaluate the proposed algorithms and
compare them to other state-of-the-art TDE algorithms. We investigate their
performance for well-spaced pulses and for overlapping pulses, and we evaluate
the estimators' performance under varying levels of measurement noise and
signal noise. Finally, \secref{sec:discussion} discusses our results and \secref{sec:conclusion} concludes the paper.

\section{Previous Work}
\label{sec:previous_work}
Prior work on the problem of sparsity in parametric dictionaries includes
\cite{Jacques2008,Ramasamy2013}, which use a gradient descent approach to
approximate solutions off the grid for a generic greedy algorithm. Another
common method is to use parabolic or polynomial interpolation on a sampled
autocorrelation function to increase the precision for sampled data
\cite{Boucher1981,Jacovitti1993,Aiordachioaie2010}.  The simplest and most often
used polynomial interpolation is fitting a parabola around the correlation peak.
In \cite{Jacovitti1993} it is proposed to use a Direct Correlator function for
parabolic interpolation:
\begin{align*}
	b_i = -\frac{\Delta}{2}\frac{\hat{R}_\vf[n+1] -
    \hat{R}_\vf[n-1]}{\hat{R}_\vf[n+1] - 2\hat{R}_\vf[n] +
    \hat{R}_\vf[n-1]} + n\Delta,
    \label{eqn:parabolic_interpolation}
\end{align*}
where $b_i$ is the translation parameter to estimate, $\Delta$ is the spacing in
time between samples of the discrete autocorrelation function:
\begin{align}
    \hat{R}_\vf[m] = \sum_{l=1}^N \vf[l]\cdot \vf[l-m],
\end{align}
and $n$ is the index of the largest absolute entry in $\hat{R}_\vf$. This
estimator is easily implemented in a greedy algorithm, where an estimate of
the discrete autocorrelation is readily available as the signal proxy.  In some
cases, it is possible to improve the estimation using different polynomial
interpolation techniques for different problems, see, e.g., the references in
\cite{Viola2005}. Interpolation-based algorithms improve the estimation
precision but suffer from interference problems if the signal components are
not orthogonal to each other.  The polynomial interpolation approach is similar
to one of the two algorithms proposed in \cite{Ekanadham2011}, one using a
first-order Taylor expansion, the other using a form of polar interpolation. The
authors show that polar interpolation outperforms Taylor expansion. In our work
we extend upon the polar interpolation approach.  In
\cite{Duarte2012,Duarte2013,Fannjiang2012}, the authors use coherence rejection
to better estimate a solution. Additionally, \cite{Duarte2013} uses polynomial
interpolation. In \cite{Fannjiang2012} the coherence rejection is implemented as
functions that inhibit coherent atoms in the recovery algorithms.  This function
is used in greedy algorithms to trim the proxy before selecting the strongest
correlating atom. Based on the coherence between a subset $S$ of atoms from the
dictionary and its complement, we can define the $\eta$-coherence band of the
index set $S$ as 
\begin{align}
    B_\eta(S) &= \bigcup_{k\in S} \{i \:|\: \mu(i,k) > \eta\},\; i \in
\{1,2,\ldots,P\}, 
    \label{eqn:bomp} 
\end{align} 
where $\mu(i,k)=|\langle\vg(b_i),\vg(b_k)\rangle|$ is the coherence between two
atoms, $\vg(b_i)$ and $\vg(b_k)$ in the dictionary $\mD$.  The authors use the
band exclusion function to avoid selecting coherent dictionary elements in
various greedy algorithms.  When applied to the Orthogonal Matching Pursuit
(OMP) algorithm, the resulting enhanced algorithm is called Band-excluded
Orthogonal Matching Pursuit (BOMP) \cite{Fannjiang2012}.

More recently, it has been shown that one can recover a
frequency-sparse signal from a random subset of its samples
using atomic norm minimization \cite{Tang2012}. The atomic norm of a signal $f$
is defined as the size of the smallest scaled convex hull of a
continuous dictionary of complex exponentials. 
Thus, the recovery procedure searches over a continuous dictionary rather than a 
discretized one. The atomic norm minimization can
be implemented as a semidefinite program (SDP), which can
be computationally expensive. In addition, this formulation
does not account for measurement noise, and it is not clear if
guarantees can be given for arbitrary measurement settings. While revised formulations have been proposed that can deal with noisy measurements at the expense of additional computation by using total variation norms~\cite{CandesFG2013} or more elaborate matrix factorizations~\cite{ChenChi2013}, the restriction to uniform low-rate (or low-count) sampling acquisition schemes appears to remain necessary. Additional efforts based on group testing~\cite{Boufounos2012} require randomized sampling schemes.

Another approach to time delay estimation is to use FFT-based methods, where the
problem is converted to a frequency estimation problem and solved using line
spectral estimation approaches such as the Multiple Signal Classification
(MUSIC) algorithm \cite{Schmidt1986} or the Estimation of Signal Parameters via
Rotational Invariance Techniques (ESPRIT) algorithm \cite{Roy1989}. This
approach exploits the fact that the dictionary matrix is cyclic. In
\cite{Saarnisaari1997}, the TDE problem is converted to an FE problem and solved 
by means of the ESPRIT algorithm. This is done by pre-multiplying the matrix
product $\mG^{-1}\mF$ on the received signal vector, $\vf = \mD\vx$:
\begin{align}
	\vy = \mG^{-1}\mF\vf = \mG^{-1}\mF\mD\vx = \mG^{-1}\mG\mF\vx = \mF\vx,
    \label{eqn:tde_music}
\end{align}
where $\mG$ is a diagonal matrix with the Fourier transform of the first column
of $\mD$ on the diagonal and zero elsewhere and $\mF$ is the DFT matrix. Because
$\mD$ is a cyclic matrix, it is diagonalized by the DFT matrix, i.e.,
$\mF\mD=\mG\mF$. Then $\vy$ contains a sum of complex exponentials and we may then use a
super-resolution algorithm to estimate the frequencies, which can be directly
mapped to delays.  However, this method has certain pitfalls. As mentioned in
\cite{LeBastard2010,Li1998} the spectrum of the pulse in $\mG$ must be nonzero
everywhere and the noise can no longer be assumed white, due to the
multiplication with the inverse of the known spectrum. The signal used in this
work spans the entire spectrum in which we sample and therefore does not suffer
from the first problem. The noise will be colored, but in our numerical
experiments this does not seem to decrease the performance much. 

A similar method has also been implemented using analog filters in
\cite{Gedalyahu2010}, which draws on a closely related estimation framework known as finite rate of innovation (FROI)~\cite{Vetterli02,Blu08}. FROI methods rely on the design of custom sampling filters, where the samples measured can be processed to obtain the desired parameter estimates; however, they have limitations similar to those of the
approach in \eqnref{eqn:tde_music}. FROI, when applied 
to TDE, relies on filters that are tailored to the Fourier transform of the signal, similarly to 
the $\mG$ matrix used in the above. These filters must be stably invertible, which 
becomes a problem if the spectrum is zero or close to zero at some frequencies.
Furthermore, these filters must also result in a coloring of the noise.  The
method in \cite{Saarnisaari1997} may also be used with CS by first
reconstructing the signal using, e.g., $\ell_1$ synthesis as in
\eqnref{eqn:synthesis} followed by estimation.

\section{Polar Interpolation}
\label{sec:polar_interpolation}
One way to remedy the discretization of the parameter space implicit in CS is to
use interpolation.  In \cite{Ekanadham2011}, a \textit{polar interpolation}
approach for translation-invariant signals has been derived. Such signals can be
written as a linear combination of shifted versions of a waveform. In a
nutshell, the interpolation procedure exploits the fact that translated versions
of a waveform form a manifold which lies on the surface of a hypersphere. Thus,
any sufficiently small segment of the manifold can be well-approximated by an
arc of a circle, and an arbitrarily-shifted waveform can be accurately approximated
by a point in one such arc connecting dictionary elements.

In this section, we first shortly define some key equations from \cite{Ekanadham2011} in \secref{sec:pi_prev}, followed by our proposed extension in \secref{sec:pi_our}.

\subsection{Previous Work on Polar Interpolation}
\label{sec:pi_prev}
Define the signal of interest as in \eqnref{eqn:signal_model}.
Then the dictionary $\mD$ from \eqnref{eqn:dictionary} samples the
translation parameter space with step size $\Delta$, and we approximate each
segment of the manifold $\{\mathcal{M}_{\vg}(b_n),\; b_n \in
[b_p-\tfrac{\Delta}{2} ,b_p+\tfrac{\Delta}{2}]\}$ by a circular arc containing the
three signals $\{\vg(b_p-\tfrac{\Delta}{2}), \vg(b_p),
\vg(b_p+\tfrac{\Delta}{2})\}$. Making use of trigonometric identities, the polar
interpolation approximates the waveform $\vg(b_n)$ using the arc containing
$b_p$, where $b_p = \llbracket b_n \rrbracket = \text{round}\left(\frac{b_n}{\Delta}\right)
\Delta$, so that $b_n = b_p + \Delta_n,\; \Delta_n \in
(-\tfrac{\Delta}{2},\tfrac{\Delta}{2})$.  Here, $b_p=\llbracket b_n\rrbracket $ signifies the
selection of the closest atom in the dictionary $b_p$ to the input parameter
$b_n$.  This arc is parametrized as follows \cite{Ekanadham2011}:
{\small \begin{align}
    \nonumber \tilde{\vg}(b_n) &= \vc(b_p) + r \cos \left(\frac{2 \Delta_n}{\Delta} \theta \right) \vu(b_p) + r \sin \left(\frac{2 \Delta_n}{\Delta} \theta \right) \vv(b_p), \\
\begin{bmatrix} \vc(b_p)^T \\ \vu(b_p)^T \\ \vv(b_p)^T \end{bmatrix} &= \begin{bmatrix} 1 & r\cos(\theta) & -r\sin(\theta) \\ 1 & r & 0 \\ 1 & r\cos(\theta) & r\sin(\theta) \end{bmatrix}^{-1} \begin{bmatrix} \vg(b_p-\tfrac{\Delta}{2})^T \\ \vg(b_p)^T \\ \vg(b_p+\tfrac{\Delta}{2})^T \end{bmatrix},
\label{eqn:approximation}
\end{align}}
where $r$ is the $\ell_2$ norm of each element of the dictionary and $\theta$ is the
angle between $\vg(b_p)$ and $\vg(b_p - \tfrac{\Delta}{2})$:
\begin{align}
	\nonumber r &= \|\vg(b_p)\|_2,\\
	\nonumber \theta &= \acos\left(\frac{\R\{\langle \vg(b_p),\vg(b_p - \tfrac{\Delta}{2}) \rangle\}}{\|\vg(b_p)\|_2 \cdot \|\vg(b_p - \tfrac{\Delta}{2})\|_2}\right)
\end{align}
for all $p \in \{1,2,\ldots,P\}$. In order to extend the above approximation to
include multiple waveforms, we introduce three dictionaries that sample the
parameter space $\llbracket \Omega_J\rrbracket  = \{\llbracket b_1 \rrbracket,
\llbracket b_2 \rrbracket, \ldots, \llbracket b_J \rrbracket\}$:
\begin{align}
    \tilde{\vf}(\Omega_J) &= \mC(\llbracket \Omega_J\rrbracket )\valpha + \mU(\llbracket \Omega_J\rrbracket )\vbeta + \mV(\llbracket \Omega_J\rrbracket )\vgamma, \nonumber \\
	\mC(\Omega_J) &= \begin{bmatrix} \vc(\llbracket b_1 \rrbracket) & \vc(\llbracket b_2 \rrbracket) & \cdots &
\vc(\llbracket b_J \rrbracket)\end{bmatrix}\in \mathbb{C}^{N\times J}, \nonumber \\
	\mU(\Omega_J) &= \begin{bmatrix} \vu(\llbracket b_1 \rrbracket) & \vu(\llbracket b_2 \rrbracket) & \cdots & \vu(\llbracket b_J \rrbracket)\end{bmatrix}\in \mathbb{C}^{N\times J}, \nonumber \\
	\label{eqn:f_approx}\mV(\Omega_J) &= \begin{bmatrix} \vv(\llbracket b_1 \rrbracket) & \vv(\llbracket b_2 \rrbracket) & \cdots & \vv(\llbracket b_J \rrbracket)\end{bmatrix}\in \mathbb{C}^{N\times J},
\end{align}
where  $\valpha$ represents the amplitude of the signal and $\vbeta$ and
$\vgamma$ controls the parameter translations. 

The three coefficient vectors, $\valpha, \vbeta$ and $\vgamma$, can be estimated
using the following constrained convex optimization problem from \cite{Ekanadham2011}, which is a variant
of the classical Basis Pursuit Denoising algorithm \cite{Chen1998}:
\begin{align}
	\label{eqn:old_convex_opt}
    (\valpha,\vbeta,\vgamma) &= \mathrm{T}(\vf, \Omega_J) \\
    \nonumber&=\argmin_{\valpha,\vbeta,\vgamma} 
    \frac{1}{2\sigma^2} \|\vf-\tilde{\vf}(\Omega_J)\|_2^2 + \lambda\|\valpha\|_1 \\
    \nonumber&\text{s.t. } \left\{\begin{matrix} \alpha_j \geq 0, \\ \sqrt{\beta_j^2 + \gamma_j^2} \leq \alpha_j r, \\ \alpha_j r\cos(\theta) \leq \beta_j \leq \alpha_j r, \end{matrix}\right\} \;\text{for } j=1,\ldots,J,
\end{align}
where $\vf$ is the received signal and $\sigma^2$ is the squared norm
of the measurement and signal noise. Here, $\lambda$ is used as a weighting factor
between sparsity and fidelity. The constraints for the optimization problem
ensure that the solution consists of points on the arcs used for approximation.
The first constraint ensures we have only nonnegative signal amplitudes. The
second enforces the trigonometric relationship among each triplet $\alpha_j$,
$\beta_j$, and $\gamma_j$. The last constraint ensures that the angle between
the solution and $\vg(b_p)$ is restricted to the interval $[-\theta,\theta]$.  It is
necessary to scale $\vbeta$ and $\vgamma$ after the optimization
problem~\cite{Ekanadham2011}:
\begin{align}
    (\beta_j, \gamma_j) \leftarrow \left(\frac{\beta_j \alpha_j r}{\sqrt{\beta_j^2+\gamma_j^2}}, \frac{\gamma_j \alpha_j r}{\sqrt{\beta_j^2+\gamma_j^2}}\right),\; \forall j. \label{eqn:normalization}
\end{align}
This is because the inequality of the second constraint should in fact be an
equality. However, the equality would violate the convexity assumption of the
optimization.  After this normalization, we obtain the signal estimate from
\eqnref{eqn:f_approx} and the frequency estimates using the one-to-one relation: 
{\small \begin{equation} 
	\label{eqn:polarint}\alpha_n \vc(b_p) + \beta_n \vu(b_p) + \gamma_n \vv(b_p)
    \approx a_n \vg\left(b_p +
    \tfrac{\Delta}{2\theta}\tan^{-1}(\tfrac{\gamma_n}{\beta_n})\right),
\end{equation}}
where the argument of $\vg(\cdot)$ is the estimate of $b_n$.  The change in
index from $j$ to $n$ is because only the $K$ absolute largest entries in
$\valpha$ and the corresponding entries in $\vbeta$ and $\vgamma$ are used for
estimation, as they represent the active atoms. The authors in
\cite{Ekanadham2011} have named this algorithm Continuous Basis Pursuit (CBP).
However, their formulation assumes non-negative real values for the amplitude
coefficients $\va$, which precludes many real-world settings. Additionally,
their choice of fidelity/sparsity trade-off in the convex optimization
formulation does not distinguish between noise and approximation error in the polar
interpolation. To address these issues, we propose an improved convex
optimization formulation in this section.

\subsection{Proposed Extension to Previous Work}
\label{sec:pi_our}
One of the contributions of this paper is an improved convex optimization
formulation of \eqnref{eqn:old_convex_opt}. To achieve this we first introduce
a metric for the approximation error, which is used together with the signal
and measurement noise $\sigma^2$ as a measure of uncertainty in the fidelity of
the solution in the optimization problem. The reason for these approximation
errors is that the fitting of a circle to the manifold is rarely perfect. This
approximation error $\delta$ is a function of the choice of waveform
$\vg(\cdot)$, spacing $\Delta$, and the translation parameter $b_n$. Let $b_n =
b_p+\Delta_n$ be an arbitrary parameter value, defined using an atom in the
dictionary $b_p$ and the translation variable $\Delta_n \in
(-\tfrac{\Delta}{2},\tfrac{\Delta}{2})$. The interpolation is based on the
assumption that the ratio between $\Delta/2$ and the arbitrary translation
variable $\Delta_n$ is equal to the ratio between $\theta$ and the angle
$\theta_n$ between $\vg(b_p)$ and $\vg(b_n)$. Define the ratio of angles as:
\begin{align*}
	\frac{\theta_n}{\theta} &= \frac{\acos\left(\R\{\langle \vg(b_p), \vg(b_n) \rangle\}\right)}{\acos\left(\R\{\langle \vg(b_p), \vg(b_p+\tfrac{\Delta}{2}) \rangle\}\right)},
\end{align*}
where we assume $r = 1$ for simplicity. Therefore, define the following bound:
\begin{align}
	\left| \frac{\acos\left(\R\{\langle \vg(b_p), \vg(b_n) \rangle\}\right)}{\acos\left(\R\{\langle \vg(b_p), \vg(b_p+\tfrac{\Delta}{2}) \rangle\}\right)} \right| \leq \left|\frac{\Delta_n}{\Delta/2}\right| + \delta \nonumber
\end{align}
This bound cannot be calculated in closed form for all classes of waveforms
$\vg(\cdot)$, but it may be numerically simulated for a given $\vg(\cdot)$ and choices of $\Delta$ and
$b_n$. Assuming that the manifold is smooth, it is possible to find the
approximation error $\delta$ as a function of $\Delta$ and $\Delta_n$ for all
possible choices of $b_p$. Then, by finding the maximum value of that function,
we obtain the worst-case bound on the interpolation error. To compute this bound
$\zeta$, we find the distance between the actual vector on the manifold and the
approximated vector, based on the value of $b$ that gives the maximum error:
{\small \begin{align}
	\nonumber\zeta &= \|\vg(\hat{b})-\tilde{\vg}(\hat{b})\|_2,\\
	\hat{b} &= \argmax_{b} \left| \frac{\acos\left(\R\{\langle \vg(b_p), \vg(b) \rangle\}\right)}{\acos\left(\R\{\langle \vg(b_p), \vg(b_p+\tfrac{\Delta}{2}) \rangle\}\right)} - \frac{\Delta_n}{\Delta/2}\right|
    \label{eqn:approximation_error}
\end{align}}
This value may then be input into the convex optimization solver.  The reason
why the error is found on the signal $\tilde{\vg}(\hat{b})$, rather than on the
parameter estimate $\hat{b}$, is because the fidelity constraint in the convex
optimization formulation is based on the function reconstruction error.

To include the approximation error and extend the optimization problem to also
allow for arbitrary complex amplitude coefficients, we reformulate the problem
formulation from \eqnref{eqn:old_convex_opt} using variable substitution:
\begin{align}
    \nonumber\valpha &= \valpha^{r,p}-\valpha^{r,n}+j(\valpha^{i,p}-\valpha^{i,n}),\; 
        \valpha \in \mathbb{C}^{1\times J}\\
    \nonumber\vbeta  &= \vbeta^{r,p}-\vbeta^{r,n}+j(\vbeta^{i,p}-\vbeta^{i,n}),\;
        \vbeta \in \mathbb{C}^{1\times J}\\
    \nonumber\vgamma &= \vgamma^{r,p}-\vgamma^{r,n}+j(\vgamma^{i,p}-\vgamma^{i,n}),\;
        \vgamma \in \mathbb{C}^{1\times J}\\
    \nonumber\vx_{\valpha} &= \begin{bmatrix} \valpha^{r,p} & \valpha^{r,n} &
        \valpha^{i,p} & \valpha^{i,n} \end{bmatrix},\; 
        \vx_{\valpha} \in (\mathbb{R}^+)^{1\times 4J}\\
    \nonumber\vx_{\vbeta} &= \begin{bmatrix} \vbeta^{r,p} & \vbeta^{r,n} &
        \vbeta^{i,p} & \vbeta^{i,n} \end{bmatrix},\; 
        \vx_{\vbeta} \in (\mathbb{R}^+)^{1\times 4J}\\
    \nonumber\vx_{\vgamma} &= \begin{bmatrix} \vgamma^{r,p} & \vgamma^{r,n} &
        \vgamma^{i,p} & \vgamma^{i,n} \end{bmatrix},\; 
        \vx_{\vgamma} \in (\mathbb{R}^+)^{1\times 4J}\\
    \nonumber\vx &= \begin{bmatrix} \vx_{\valpha} & \vx_{\vbeta} & \vx_{\vgamma} \end{bmatrix}^T,\\
\mE(\Omega_J) & = \left[
        \mC(\Omega_J)~~-\mC(\Omega_J)~~j\mC(\Omega_J)~~-j\mC(\Omega_J)\right. \nonumber\\
        & \quad~~
        \left. \mU(\Omega_J)~~-\mU(\Omega_J)~~\cdots~~-j\mV(\Omega_J)
       \right].     \label{eqn:advanced_variables} 
\end{align}
In the above, we use the shorthand notation $\mathbb{R}^+ = \{x \in \mathbb{R} : x \ge 0\}$. Each optimization variable from the previous optimization problem is expanded into four: 1) the real positive part, 2) the real negative part, 3) the imaginary positive part and 4) the imaginary negative part. These are again used in a new optimization variable $\vx$, which is real and nonnegative.
The resulting convex optimization problem is written as:
{\small \begin{align}
    \nonumber \vx & = \mathrm{T}(\vy, \mA, \Omega_J) \\
    &=  \arg \min_{\vx,\vt} \|\vy-\mA\mE(\Omega_J)\vx\|_2^2 + 
        \frac{\lambda}{2(\sigma^2 + \zeta)}\|\vt\|_1 \label{eqn:advanced_convex_optimization} \\
    \nonumber\text{s.t. } 
    &\left\{\begin{matrix} 
        \sqrt{\vx_{\vbeta}(j)^2 + \vx_{\vgamma}(j)^2} \leq \vx_{\valpha}(j)r, \\
        \vx_{\valpha}(j)r\cos(\theta) \leq \vx_{\vbeta}(j) \leq \vx_{\valpha}(j)r,
    \end{matrix}\right\} \;\text{, } j=1,\ldots,4J, \nonumber \\
    &\left\{\begin{matrix} 
        \vt(j) \geq \sqrt{\valpha^{r,p}(j)^2 + \valpha^{r,n}(j)^2 + 
            \valpha^{i,p}(j)^2 + \valpha^{i,n}(j)^2}
    \end{matrix}\right\} \;\text{, } \nonumber\\
    &  j=1,\ldots,J, \nonumber
\end{align}}
Here we have included the CS measurement matrix $\mA$, which was not
part of the work in \cite{Ekanadham2011}. Also introduced is an auxiliary 
optimization variable $\vt$, which is used in a mixed $\ell_1-\ell_2$ norm to 
control the sparsity and the magnitude of individual components of the solution $\vx$.
This formulation allows for both complex and negative amplitudes; 
when applied with all parameter values used in the
dictionary $\mD$, we refer to it as Complex Continuous Basis Pursuit (CCBP):
\begin{align}
    (\vx) &= \mathrm{T}_\text{CCBP}(\vy, \mA, \Omega_{CCBP}),
	\label{eqn:ccbp}
\end{align}
where $\Omega_{CCBP} = \{b_1, b_2, \ldots, b_P\}$ is the set of all translation
parameters that appear in the dictionary for the application of interest.
Parameter estimates are then obtained using Eqns.
(\ref{eqn:normalization}--\ref{eqn:polarint}). CCBP has a high computational
complexity; it operates on matrices of size $M\times 12N$, whereas other CS algorithms
operate on matrices of size $M \times N$. However, its interpolation step has one
important advantage: translation-invariance and interpolation enables CCBP to
reconstruct arbitrary translation invariant sparse signals while requiring only
a small subset of the $N$ parameters to be contained in the corresponding
dictionary.  This makes it possible to incorporate the convex optimization
solver into a greedy algorithm that quickly finds a rough estimate, which is
then improved upon by a convex optimization solver.

\section{Interpolating Band-excluded Orthogonal Matching Pursuit}
\label{sec:ibomp}
To be able to leverage both the accuracy of the convex optimization solvers and
the speed of a greedy algorithm, we propose a greedy algorithm that may
improve upon its initial estimate using the convex optimization in
(\ref{eqn:advanced_convex_optimization}). In \cite{Fyhn2013} it is shown how
the Subspace Pursuit algorithm \cite{Dai2009} may be utilized for this purpose.
However, in that work the frequencies to estimate are well separated, whereas
in this work, we also evaluate the algorithms for overlapping pulses with the
band exclusion function disabled. In that case the Subspace Pursuit algorithm
may pick an incorrect dictionary element that is coherent with a strong signal
component rather than the correct dictionary element for a weak signal
component. This happens because the Subspace Pursuit algorithm attempts to find
all the pulses in the signal in one iteration. Instead, we utilize the BOMP
algorithm with interpolation, termed Interpolating Band-excluded Orthogonal
Matching Pursuit (IBOMP). This is a greedy algorithm with an optional convex
optimization problem. The algorithm improves upon the BOMP algorithm by using
interpolation in each iteration to enhance the estimate of the translation
parameter and residual. The IBOMP algorithm is shown in \algoref{algo:ibomp}. First, the
best correlating atom index $i_n$ is found by generating a proxy for the sparse
signal. This proxy is trimmed based on the band exclusion function $B_\eta(S)$,
as defined in \eqnref{eqn:bomp}, where $S$ is the current support estimate. 
The selected atom, $i_n$, is then input to an
interpolation function, $\mathrm{T}(\cdot)$. This function outputs an estimated
translation parameter, which is used to create a new atom for a signal
dictionary, $\mB$, by using the original parametric signal model from \eqnref{eqn:signal_model}. This new
signal dictionary is used to find the basis coefficients $\va$ using least
squares.  Then, a new residual is calculated and $n$ and $S$ are updated. This
loop runs $K$ times, i.e., once for each pulse in the signal. After the greedy
algorithm is done the estimates may be improved upon by running the CCBP
algorithm on a limited parameter set based on the current parameter estimates.
When exiting the loop the estimates found by the greedy algorithm are put into
a new set, $\Omega$, together with $\xi$ adjacent indices on each side. This is necessary
because the parameter values generating $\vy$ may not be sufficiently
incoherent and may therefore skew the peaks of the proxy estimate. Therefore,
as a precaution, we include the closest neighbors on each side. The set
$\Omega$ is input to the convex optimization in
(\ref{eqn:advanced_convex_optimization}) along with the measurement matrix and
the received signal. The output $\vx$ from the CCBP algorithm is then split back into $\valpha$, $\vbeta$ and $\vgamma$ values that are used to generate new
estimates of the reconstructed signal $\tilde{\vf}$ and the parameter vector
$\tilde{\vb}$.
\begin{algorithm}[bt]
    \caption{Interpolating Band-excluded Orthogonal Matching Pursuit (IBOMP)}
    \label{algo:ibomp}
    \begin{algorithmic}
        \STATE \textbf{INPUTS:} Compressed signal $\vy$, interpolation function
        $\mathrm{T}(\cdot)$, band exclusion parameter $\eta$, dictionary $\mD$, measurement matrix $\mA$ and
        number of adjacent indicies to include in the CCBP algorithm $\xi$.
        \STATE \textbf{OUTPUTS:} Reconstructed signal $\tilde{\vf}$ and parameter estimates $\tilde{\vb}$.
        \STATE Initialize: $\vy_{res}=\vy$, $\mB=\emptyset$, $n=1$ and $S^0=\emptyset$.
        \WHILE{$n \leq K$}
                \STATE $i_n = \arg\max_i |\langle\vy_{res},\mA\mD_i\rangle|,\:i \not\in B_\eta(S^{n-1})$
                \STATE $\hat{b}_n = \mathrm{T}(\vy_{res}, \mA, i_n)$
                \STATE Include sampled version of $\vg(\hat{b}_n)$ as new atom in $\mB$
                \STATE $\va = (\mA\mB)^\dagger \vy$
                \STATE $\vy_{res}=\vy-\mA\mB\va$
                \STATE $S^n = S^{n-1} \cup \{i_n\}$
                \STATE $n = n+1$
        \ENDWHILE
	    \STATE $\Omega = \cup\{\Delta(s-\xi),\ldots,\Delta(s+\xi) | s \in S^n\}$
        \STATE Use $\mathrm{T}(\vy, \mA, \Omega)$ from
        \eqnref{eqn:advanced_convex_optimization} to obtain $\vx$
        \STATE Obtain $\tilde{\vf}$ and $\tilde{\vb}$ by extracting $\valpha$, $\vbeta$ and $\vgamma$ from $\vx$ (\ref{eqn:advanced_variables}) and using (\ref{eqn:polarint}) and (\ref{eqn:f_approx})
    \end{algorithmic}
\end{algorithm} 

In this work we use two interpolation functions: parabolic interpolation and
polar interpolation.

\begin{itemize}
\item \textbf{Parabolic Interpolation Function}:
We define the parabolic interpolation function based on
\eqnref{eqn:parabolic_interpolation} as follows:
{\small \begin{align}
    \mathrm{T}_{\text{Pa}}(\vy_{res}, \mA, i_n) =
    -\frac{\Delta}{2}\frac{\hat{R}[i_n+1] - \hat{R}[i_n-1]}{\hat{R}[i_n+1] -
    2\hat{R}[i_n] + \hat{R}[i_n-1]} + i_n\Delta,
    \label{eqn:ibomp_parabolic_interpolation}
\end{align}}
where $\hat{R}[m]$ is defined as:
\begin{align}
    \hat{R}[m] = \sum_{l=1}^N \vy_{res}[l]\cdot \mA\vg[l-m], \nonumber
\end{align}
In the IBOMP algorithm there is no reason to calculate the $\hat{R}[m]$ function
as it is identical to the proxy in the greedy algorithm.
\item \textbf{Polar Interpolation Function}:
The polar interpolation function is based on \eqnref{eqn:approximation}. We
reformulate those equations to a linear least squares problem:
\begin{align}
    \vy_{res,n} &\approx \mA  
    \begin{bmatrix} 
        \vg(b_p-\frac{\Delta}{2}) & \vg(b_p) & \vg(b_p+\frac{\Delta}{2}) 
    \end{bmatrix} \times \nonumber \\
    &\quad \left(\begin{bmatrix} 
        1 & r\cos(\theta) & -r\sin(\theta) \\ 1 & r & 0 \\ 1 & r\cos(\theta) & r\sin(\theta) 
    \end{bmatrix}^{-1}\right)^T \vx,\;     \label{eqn:ibomp_polar_interpolation} \\
    \vx &= \begin{bmatrix} 
        a_i \\ a_i r\cos\left(\frac{2\Delta_n\theta}{\Delta}\right) \\ 
        a_i r\sin\left(\frac{2\Delta_n\theta}{\Delta}\right) 
    \end{bmatrix}. \nonumber
\end{align}
In this formula, a rotation matrix rotates the three $\vg$ vectors to form a
new, general basis for the circle arc and $\vx$ scales the vectors in that basis
to estimate the received signal. Given a signal or residual $\vy_{res,n}$ and
the atom $\vg(b_p)$ in the dictionary that correlates the strongest with the
residual, we may solve \eqnref{eqn:ibomp_polar_interpolation} as a linear
least squares problem with $\vx$ as the unknown. From the estimate
$\hat{\vx}=[\hat{x}_1, \hat{x}_2, \hat{x}_3]^T$, we may obtain an estimate of
$b_n=b_p+\Delta_n$ as:
\begin{align}
    b_n=b_p+\arctan\left(\frac{\hat{x}_3}{\hat{x}_2}\right)\frac{\Delta}{2\theta}. 
    \label{eqn:poibomp}
\end{align}
We term the interpolation function in \eqnref{eqn:ibomp_polar_interpolation}
$\mathrm{T}_{\text{Po}}(\vy_{res}, \mA, i_n)$, where $i_n$ is the index in the
dictionary for $\vg(b_p)$.
\end{itemize}

The IBOMP algorithm finds one estimate of a function component using either parabolic
interpolation as in \eqnref{eqn:ibomp_parabolic_interpolation} or polar
interpolation as in \eqnref{eqn:ibomp_polar_interpolation} and then removes that
estimated waveform from the residual $\vy_{res,n}$, after which it continues to
work on the residual. After the greedy algorithm has found a number of promising
estimates, we may improve upon these with the CCBP algorithm. Another solution
would be to use CCBP in each iteration of the BOMP algorithm. However, this
increases the computational complexity and in our experiments we have not found
that this improves performance.

For the band exclusion function, we set $\eta=0$ if we know that the pulses are
well spaced (i.e., orthogonal). In that case, the band exclusion does not inhibit
two pulses from interfering, but inhibits the algorithm from finding the same
pulse again due to a large remaining residual.  Otherwise, if we are only
interested in identifying pulses with a given spacing, we may adjust $\eta$ to
reflect this. If we cannot make any assumption to the spacing, we set $\eta=1$.

\section{Numerical Experiments}
\label{sec:numerical_experiments}
To evaluate the proposed algorithms, we first must find good parameter values
for the convex optimization problem. The two parameters $\zeta$ and $\lambda$
signify approximation error and sparsity trade-off, respectively. This analysis
shows why the FE problem is more complex than the TDE problem when assuming
both positive and negative complex amplitude coefficients. The analysis is
followed by experiments for the TDE problem that evaluate the proposed
algorithms in different scenarios. We investigate their performance for
well-spaced pulses and for overlapping pulses and we investigate the
performance when the signal experiences signal noise instead of measurement
noise. All the code along with the results and figures in
this paper is available at \url{www.sparsesampling.com/cpe} following the
principle of Reproducible Research \cite{Vandewalle2009}.

Before explaining the experiments further, we define the two types of signals and
the dictionaries that are used in all the following experiments. For both types,
the general signal model is as defined in \eqnref{eqn:signal_model}.

For the TDE numerical experiments, we let the pulse model $g(t)$ be a chirp
signal defined as
\begin{align}
        g(t, b_n) &= \frac{1}{\sqrt{\mathcal{E}_g}} \cdot e^{j2\pi (f_0+\frac{\Delta
        f}{2T}(t-b_n))(t-b_n)} \cdot p(t-b_n), \nonumber \\
        p(t) &= \left\{\begin{matrix} \frac{T}{2}(1 + \cos(2\pi t/T)), & t \in (0,T) \\ 0, & \text{otherwise} \end{matrix}\right., \nonumber
\end{align}
where $f_0=1$MHz is the center frequency, $\Delta f=40$MHz is the sweeped frequency
, and $T=1\mu$s is the duration of the chirp in time. The chirp is limited in time
by a raised cosine pulse and normalized to unit energy.  We generate a sampled
time signal, $\vg(b_n)$ by sampling the pulse function:
\begin{align*}
    \vg(b_n) & = 
    \begin{bmatrix} g_1(b_n) & g_2(b_n) & \cdots & g_N(b_n) \end{bmatrix},\\
    g_i(b_n) &= g((i-1)T_s, b_n)
\end{align*}
Here, $T_s$ is the sampling period. We sample the signal at $50$MHz, since the
corresponding bandwidth of the signal contains more than $99\%$ of its energy.
For each signal we take $N=500$ samples. The dictionary $\mD$ for the TDE
problem is a circulant matrix with shifted versions of $\vg(b_n)$:
\begin{align*}
\nonumber\mD_{TDE}  &= \begin{bmatrix}  \vg(b_1)    & \vg(b_2)  & \cdots & \vg(b_N) \end{bmatrix} \\
                    &= \begin{bmatrix}  g[0]        & g[N-1]    & \cdots & g[1] \\
                                        g[1]        & g[0]      & \ddots & g[2] \\
                                        \vdots      & \vdots    & \ddots & \vdots \\
                                        g[N-1]      & g[N-2]    & \cdots & g[0] \end{bmatrix},
\end{align*}
where $\vg(0)=\begin{bmatrix} g[0] & g[1] & \cdots & g[N]\end{bmatrix}^T$. This
means that the spacing between atoms in this dictionary is equal to the sampling
rate, $T_s$.

For the FE numerical experiments we generate frequency-sparse signals of length
$N=100$ containing $K$ complex sinusoids with frequencies selected uniformly at
random. The continuous signal function then becomes:
\begin{align*}
    g(t, b_n) = \frac{1}{\sqrt{N}} \exp\left(j2\pi b_n t/N\right).
\end{align*}
The basic dictionary for this signal is a DFT matrix with spacing $1$Hz between
atoms.

\subsection{Analysis of the $\zeta$ and $\lambda$ Parameters}
We first investigate the approximation error parameter, $\zeta$. We have
conducted numerical experiments on the bound in
\eqnref{eqn:approximation_error}. These experiments are conducted for both the
TDE and FE problem, to show that the approximation error is problem-specific.

\begin{figure}
        \centering
        \newlength\figureheight
        \newlength\figurewidth
        \setlength\figureheight{4.5cm}
        \setlength\figurewidth{0.4\textwidth}
%
%
%
\begin{tikzpicture}

\begin{axis}[%
width=\figurewidth,
height=\figureheight,
scale only axis,
xmin=0,
xmax=30,
xlabel={c},
ymode=log,
ymin=1e-08,
ymax=100,
yminorticks=true,
ylabel={$\zeta$},
legend style={draw=black,fill=white,legend cell align=left}
]
\addplot [
color=blue,
solid,
mark=x,
mark options={solid}
]
table[row sep=crcr]{
1 0.00768094289262555\\
2 0.000979490169288359\\
3 0.000291312855263971\\
4 0.000123067599810892\\
5 6.30554644647796e-05\\
6 3.65073159958558e-05\\
7 2.29982655018212e-05\\
8 1.54118805333516e-05\\
9 1.08274906711198e-05\\
10 7.89559827568691e-06\\
11 5.93390186508093e-06\\
12 4.57208144910222e-06\\
13 3.59727992323484e-06\\
14 2.88121059834712e-06\\
15 2.34341383729813e-06\\
16 1.93168243485169e-06\\
17 1.61113292046776e-06\\
18 1.35785322301165e-06\\
19 1.15507887619128e-06\\
20 9.90819520001981e-07\\
21 8.56345097890242e-07\\
22 7.45196707234017e-07\\
23 6.52526065856286e-07\\
24 5.74645753148287e-07\\
25 5.08717418613982e-07\\
26 4.5253215943215e-07\\
27 4.04353481699745e-07\\
28 3.62803517430115e-07\\
29 3.26779485461679e-07\\
30 2.95391686365212e-07\\
};
\addlegendentry{Polar TDE};

\addplot [
color=blue,
dashed,
mark=o,
mark options={solid}
]
table[row sep=crcr]{
1 0.357542275889494\\
2 0.180747253942922\\
3 0.120744447119712\\
4 0.0906231121283518\\
5 0.0725224937396093\\
6 0.0604462814798271\\
7 0.051816717537764\\
8 0.0453428194071431\\
9 0.0403066735271452\\
10 0.0362772584776265\\
11 0.0329801682547812\\
12 0.0302324082168588\\
13 0.0279072602756082\\
14 0.0259141954165025\\
15 0.0241868164704473\\
16 0.0226753200483349\\
17 0.0213416177919702\\
18 0.0201560832422196\\
19 0.0190953256578358\\
20 0.0181406314833425\\
21 0.0172768509887816\\
22 0.0164915884774355\\
23 0.0157746036959349\\
24 0.0151173628608899\\
25 0.0145126974188151\\
26 0.0139545415400832\\
27 0.0134377279381831\\
28 0.0129578274355558\\
29 0.0125110217180835\\
30 0.0120940015355057\\
};
\addlegendentry{BOMP TDE};

\addplot [
color=red,
solid,
mark=x,
mark options={solid}
]
table[row sep=crcr]{
1 0.406748435299452\\
2 0.0529203957185506\\
3 0.0156830302826763\\
4 0.00661409508933553\\
5 0.00338568967509537\\
6 0.00195905192196759\\
7 0.00123358382206827\\
8 0.000826357490289245\\
9 0.000580353962666428\\
10 0.000423065686107487\\
11 0.000317848566568031\\
12 0.000244820226854321\\
13 0.000192555211147538\\
14 0.000154168866869611\\
15 0.000125343936027669\\
16 0.000103279501790857\\
17 8.61042931794095e-05\\
18 7.25357231418792e-05\\
19 6.16746748817856e-05\\
20 5.28781381402697e-05\\
21 4.56779852354406e-05\\
22 3.97279026290807e-05\\
23 3.47679571209074e-05\\
24 3.06004687286276e-05\\
25 2.70732880019694e-05\\
26 2.40680163462912e-05\\
27 2.14915845269973e-05\\
28 1.92701495862839e-05\\
29 1.73446161509267e-05\\
30 1.56673114792691e-05\\
};
\addlegendentry{Polar FE};

\addplot [
color=red,
dashed,
mark=o,
mark options={solid}
]
table[row sep=crcr]{
1 1.40712472794703\\
2 0.846632526726489\\
3 0.5839763940066\\
4 0.443224200053754\\
5 0.356535773300927\\
6 0.298002181636685\\
7 0.255890921108563\\
8 0.224166418479886\\
9 0.199418765162075\\
10 0.179579778943746\\
11 0.163323582271385\\
12 0.149761572652691\\
13 0.138276150027113\\
14 0.12842485287579\\
15 0.119882453018318\\
16 0.11240457707428\\
17 0.105804071086227\\
18 0.0999351918076096\\
19 0.0946827612047734\\
20 0.0899545577684486\\
21 0.0856758716484923\\
22 0.0817855385373796\\
23 0.0782330042672644\\
24 0.074976120721528\\
25 0.0719794690551089\\
26 0.0692130687404375\\
27 0.0666513727243432\\
28 0.0642724773756523\\
29 0.0620574955167768\\
30 0.059990054581334\\
};
\addlegendentry{BOMP FE};

\end{axis}
\end{tikzpicture}%
      	\caption{Polar estimation approximation error analysis: the figure shows the approximation error parameter $\zeta$ as a function of the parametric dictionary redundancy factor $c$.}
        \label{fig:approximation_error_analysis}
\end{figure}
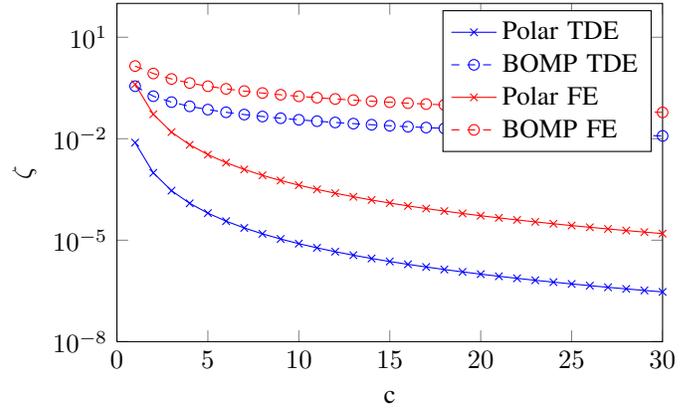

The approximation error from \eqnref{eqn:approximation_error} depends on the
specific signal model and the dictionary spacing $\Delta$. For each of the two
signal models, we have performed numerical experiments for a range of spacings.
For the TDE problem, the spacing is defined as $\Delta = \frac{T_s}{c}$, where
$c$ is called the redundancy factor and is used as the experiment variable. For
the FE problem, the spacing is defined as $\Delta = \frac{1}{c}$. In each
experiment, we pick a center atom in the dictionary and uniformly sample the
parameter space around that atom using $100$ samples. Each sample constitutes a
parameter value $b$ to input into the equations in
\eqnref{eqn:approximation_error}. The result of the experiment is shown in
\figref{fig:approximation_error_analysis}.  In the figure, the approximation
error is compared to the maximum approximation noise from BOMP, i.e., the
approximation error when a parameter lies exactly in between two atoms in the
dictionary.
As can be seen, the FE signals suffer from a higher approximation
noise than the TDE case. For the TDE problem, good performance is achievable
without any redundancy in the dictionary, i.e., for $c=1$, whereas for FE a
higher redundancy factor is needed. This, however, increases computation time
significantly and introduces coherence. 

When the coherence of the dictionary increases, it becomes less likely to find a
unique, sparse solution to the problem. This is because of the coherence of the
redundant dictionary and the looseness of the fidelity constraint in
\eqnref{eqn:advanced_convex_optimization}, due to the approximation error.
This is best illustrated using the Spark of the dictionary. Given a matrix
$\mD$ we define $\sigma = \text{Spark}(\mD)$ as the smallest possible number
such that there exists a subgroup of $\sigma$ columns from $\mD$ that are
linearly dependent \cite{Donoho2003}. The Spark is computationally heavy to
compute, but an upper bound can be found \cite{Donoho2003}. Define a sequence of
optimization problems, $i=1,\ldots,N$:
\begin{align*}
    \tilde{\vx}_i^0 &= \min_{\vx \in \mathbb{C}^N} \|\vx\|_0 ~~\textrm{s.t.}~~ \mD\vx = \bar{0},\;
    \vx_i = 1, \\
    \text{Spark}(\mD) &= \min_{1 \leq i \leq N} \|\tilde{\vx}_i^0\|_0
\end{align*}
The optimization problem is however not computationally feasible due to
the $\ell_0$ term. Instead, we use a $\ell_1$ norm, which is solvable in
polynomial time using standard solvers. Because $\|\tilde{\vx}_i^0\|_0 \leq
\|\tilde{\vx}_i^1\|_0$, we obtain the upper bound on the Spark:
\begin{align*}
    \tilde{\vx}_i^1 &= \min_{\vx \in \mathbb{C}^N} \|\vx\|_1 ~~\textrm{s.t.}~~ \mD\vx = \bar{0},\;
    \vx_i = 1,\\
    \text{Spark}(\mD) &\leq \min_{1 \leq i \leq N} \|\tilde{\vx}_i^1\|_0
\end{align*}
Using the two dictionaries defined for the TDE and FE problems, we have found
this upper bound on the Spark. For the TDE problem
$\text{Spark}(\mD_{TDE})\leq N$, because all the columns are linearly
independent. There is no redundancy and the matrix has full rank. Hence,
coherence is not a problem in the TDE case. For the FE problem with $c=5$ we
have $\text{Spark}(\mD_{\text{FE}})\leq101$.  This problem seems to
contradict the results from \cite{Fyhn2013}, where polar interpolation works
well for the FE problem. However, in that work the amplitude coefficients are
real and non-negative. If we find the spark with those assumptions, i.e., solve
the following optimization problem:
\begin{align*}
    \tilde{\vx}_i^1 &= \min_{\vx \in \mathbb{R}^N} \|\vx\|_1 ~~\textrm{s.t.}~~ \mD\vx = \bar{0},\;
    \vx_i = 1,\; \vx \geq \bar{0},\\
    \text{Spark}_{+}(\mD) &\leq \min_{1 \leq i \leq N} \|\tilde{\vx}_i^1\|_0,
\end{align*}
the upper bound becomes $\text{Spark}_{+}(\mD_{\text{FE}})\leq N$ even though the matrix
$\mD$ does not have full rank.

The first result shows that polar interpolation is easier to apply to the TDE
problem, at least with the signal model chosen for this work, than to the FE
problem. It does not mean that polar interpolation cannot be applied to the FE
problem, but it will require a different convex optimization formulation in
which the constraints are tightened further, to shrink the solution set. This
may be possible in the case where the problem allows for some specific
assumptions, e.g., some symmetry in the spectrum which can be formulated as
constraints in the optimization problem.

\begin{figure*}
    \centering
    \setlength\figureheight{4.5cm}
    \setlength\figurewidth{0.3\textwidth}
    \subfloat{
%
%
%
\begin{tikzpicture}

\begin{axis}[%
width=\figurewidth,
height=\figureheight,
view={0}{90},
scale only axis,
xmin=0.2,
xmax=1,
xlabel={$\kappa$},
ymode=log,
ymin=1e-06,
ymax=1000000,
yminorticks=true,
ylabel={$\lambda$},
zmin=-10,
zmax=0,
zlabel={b-MSE},
title={Number of runs:36},
axis x line*=bottom,
axis y line*=left,
axis z line*=left,
colormap/blackwhite,
colorbar,
point meta min=-8.01255867554213,
point meta max=-4.41582264728611
]

\addplot3[%
surf,
colormap/blackwhite,
shader=faceted,
draw=black,
mesh/rows=5]
table[row sep=crcr,header=false] {
0.2 1e-06 -7.74338414982521\\
0.2 1e-05 -7.74379135359792\\
0.2 0.0001 -7.74363590890328\\
0.2 0.001 -7.74308534570894\\
0.2 0.01 -7.76027045976657\\
0.2 0.1 -7.78569021948481\\
0.2 1 -7.6793389015243\\
0.2 10 -6.15901753863729\\
0.2 100 -4.49095164858068\\
0.2 1000 -4.43728817180872\\
0.2 10000 -4.43728817180872\\
0.2 100000 -4.43728817180872\\
0.2 1000000 -4.43728817180872\\
0.4 1e-06 -7.55190388023453\\
0.4 1e-05 -7.61162998464862\\
0.4 0.0001 -7.68968039923491\\
0.4 0.001 -7.70566807659445\\
0.4 0.01 -7.7882061885029\\
0.4 0.1 -7.9292082168311\\
0.4 1 -7.8244683695657\\
0.4 10 -6.61680258339291\\
0.4 100 -5.10681805839383\\
0.4 1000 -4.63546933056682\\
0.4 10000 -4.62397021026286\\
0.4 100000 -4.62397021026286\\
0.4 1000000 -4.62397021026286\\
0.6 1e-06 -7.10785622610915\\
0.6 1e-05 -7.24619604356618\\
0.6 0.0001 -7.34616191267683\\
0.6 0.001 -7.46868148039928\\
0.6 0.01 -7.68120524442572\\
0.6 0.1 -7.86511136925137\\
0.6 1 -7.75773291232972\\
0.6 10 -6.68972871465264\\
0.6 100 -4.91049289095375\\
0.6 1000 -4.51817654864568\\
0.6 10000 -4.47603470270511\\
0.6 100000 -4.47603470270511\\
0.6 1000000 -4.47603470270511\\
0.8 1e-06 -6.99992767336237\\
0.8 1e-05 -6.99983296796492\\
0.8 0.0001 -7.01340631721253\\
0.8 0.001 -7.27392437053029\\
0.8 0.01 -7.58367834440525\\
0.8 0.1 -7.89509901037966\\
0.8 1 -7.76672406975791\\
0.8 10 -7.42097640059625\\
0.8 100 -5.56818859572661\\
0.8 1000 -4.51334146623213\\
0.8 10000 -4.59408379925141\\
0.8 100000 -4.59408379925141\\
0.8 1000000 -4.59408379925141\\
1 1e-06 -6.74920124813022\\
1 1e-05 -6.9883397260619\\
1 0.0001 -7.03991634899484\\
1 0.001 -7.30377638935497\\
1 0.01 -7.63638886693145\\
1 0.1 -8.01255867554213\\
1 1 -7.83485312201632\\
1 10 -7.80216211746928\\
1 100 -6.1763562975988\\
1 1000 -4.47853861929834\\
1 10000 -4.41582264728611\\
1 100000 -4.41582264728611\\
1 1000000 -4.41582264728611\\
};
\end{axis}
\end{tikzpicture}%
}
    \subfloat{
%
%
%
\begin{tikzpicture}

\begin{axis}[%
width=\figurewidth,
height=\figureheight,
view={0}{90},
scale only axis,
xmin=0,
xmax=30,
xlabel={SNR [dB]},
ymode=log,
ymin=1e-06,
ymax=1000000,
yminorticks=true,
ylabel={$\lambda$},
zmin=-10,
zmax=10,
zlabel={b-MSE},
title={Number of runs:42},
axis x line*=bottom,
axis y line*=left,
axis z line*=left,
colormap/blackwhite,
colorbar,
point meta min=-7.773419568525,
point meta max=1.49939839671944
]

\addplot3[%
surf,
colormap/blackwhite,
shader=faceted,
draw=black,
mesh/rows=4]
table[row sep=crcr,header=false] {
0 1e-06 1.49939839671944\\
0 1e-05 1.47979229629697\\
0 0.0001 1.28733296399126\\
0 0.001 1.18612867426088\\
0 0.01 -4.40595520923103\\
0 0.1 -5.04144135834043\\
0 1 -5.35468133562959\\
0 10 -4.53534615368699\\
0 100 -4.50304300383869\\
0 1000 -4.50304300383869\\
0 10000 -4.50304300383869\\
0 100000 -4.50304300383869\\
0 1000000 -4.50304300383869\\
10 1e-06 1.38870459481583\\
10 1e-05 1.46299807514553\\
10 0.0001 1.46335436651467\\
10 0.001 0.340392047875526\\
10 0.01 -5.3328716024572\\
10 0.1 -6.10959914067178\\
10 1 -6.08009115591787\\
10 10 -4.81312245092641\\
10 100 -4.50135833942596\\
10 1000 -4.50135833942596\\
10 10000 -4.50135833942596\\
10 100000 -4.50135833942596\\
10 1000000 -4.50135833942596\\
20 1e-06 1.46427087635999\\
20 1e-05 1.47350794009434\\
20 0.0001 1.14298927620257\\
20 0.001 -4.34631561924002\\
20 0.01 -6.07247384309845\\
20 0.1 -6.60808958973162\\
20 1 -6.83561892134798\\
20 10 -5.63437903180784\\
20 100 -4.44892872895576\\
20 1000 -4.4410395655985\\
20 10000 -4.4410395655985\\
20 100000 -4.4410395655985\\
20 1000000 -4.4410395655985\\
30 1e-06 1.4551276971393\\
30 1e-05 1.33931759843401\\
30 0.0001 -1.21403756869588\\
30 0.001 -6.0107226848756\\
30 0.01 -7.03097144352543\\
30 0.1 -7.58612479089156\\
30 1 -7.773419568525\\
30 10 -6.63888903190279\\
30 100 -4.8885249323333\\
30 1000 -4.64792628189423\\
30 10000 -4.64792628189423\\
30 100000 -4.64792628189423\\
30 1000000 -4.64792628189423\\
};
\end{axis}
\end{tikzpicture}%
}
    \caption{$\lambda$ analysis for CCBP. Left figure is with SNR $1000$ and
	right figure is with $\kappa=1$. The z-axis is the mean squared error of 
	the parameter estimate $\vb$-MSE. The scaling on the z-axis is microseconds squared 
	on a logarithmic scale.}
    \label{fig:lambda_analysis}
\end{figure*}
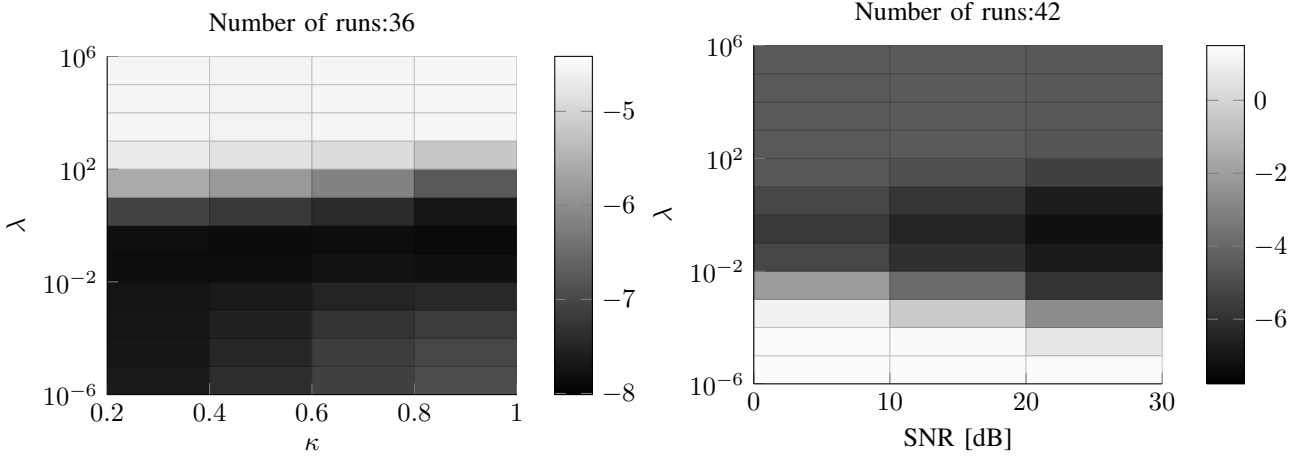

In the following we limit our focus to the TDE problem and show the estimator
performance in different scenarios. First, however, we must investigate what the
other optimization variable, $\lambda$, shall be. This is no trivial task as its
optimal value changes depending on the function $g(t)$, the subsampling ratio
$\kappa$, the SNR, etc.  To visualize this and to find a good candidate for
$\lambda$ for later experiments, we have evaluated different choices of $\lambda$
for the signal $\vg(\cdot)$ used in our TDE problem, while also varying $\kappa$ and the SNR. The estimator
performance is evaluated in terms of the mean squared error (MSE) on the $\vb$
parameter, termed the $\vb$-MSE. This corresponds to the sample variance of the
estimators and is a measure of estimator precision. We perform Monte Carlo
experiments to get an average result on the error. In each experiment, we
generate a time signal with one pulse ($K=1$) by sampling the signal function in
\eqnref{eqn:signal_model}. The real and imaginary part of the amplitude
coefficient $a$ are drawn from a uniform distribution between $1$ and $10$. As
shown in \figref{fig:approximation_error_analysis} there is no need for a
redundant dictionary matrix for the TDE problem, so we use $c=1$, i.e., the
dictionary has size $500\times 500$. The results are shown in
\figref{fig:lambda_analysis}.  The colorbar signifies the $\vb$-MSE in
microseconds squared on a logarithmic scale. As $\lambda$ increases the
$\ell_1$-norm of the solution vector $\vx$ decreases and eventually becomes the
zero-vector. When this happens the CCBP algorithm falls back to the BOMP
algorithm. As can be seen, $\lambda=1$ is a good choice for TDE estimation.

\subsection{Performance Evaluation of the Estimators}
Now the optimization variables for the TDE problem have been chosen and the next
experiment shall evaluate the estimation performance of the three proposed
algorithms versus other TDE algorithms when CS is applied. We evaluate the
estimators in three different scenarios:
\begin{itemize}
    \item Case A: Experiments for well-spaced pulses with and without
        measurement noise.
    \item Case B: Experiments for overlapping pulses with and without 
        measurement noise.
    \item Case C: Experiments for overlapping pulses with
        signal noise.
\end{itemize}
The two first cases evaluate how much the signal may be subsampled using CS and
still attain good estimation precision. The last case evaluates the effect of
noise folding, when the noise is added before the measurement matrix $\mA$ is
multiplied on.

For all the experiments we use a Random Demodulator CS measurement matrix
\cite{Tropp2010}, $\mPsi\in\{-1,0,1\}^{M\times N}$. We set $M=\kappa N$, where
$\kappa\in[0,0.5)$ is the CS subsampling rate. We evaluate the performance of
the three estimators by computing the translation parameter mean squared
error ($\vb$-MSE) between the true value of the time delay and the estimated
value. Each point in the plot is the result of more than $100$ Monte Carlo
experiments. The algorithms we evaluate are as follows:
\begin{itemize}
    \item \textbf{BOMP} - a greedy algorithm proposed in \cite{Fannjiang2012}
        with no interpolation,
    \item \textbf{BISP} - similar to BOMP followed by CCBP, but using Subspace Pursuit \cite{Dai2009} rather than OMP, proposed in \cite{Fyhn2013},
    \item \textbf{TDE-MUSIC} - an algorithm that reconstructs the received
        signal using \eqnref{eqn:synthesis} after which the problem is converted
        to a frequency estimation problem that is solved using the MUSIC
        algorithm, as explained in \eqnref{eqn:tde_music},
    \item \textbf{PaIBOMP} - BOMP with parabolic interpolation,
    \item \textbf{CCBP} - The CCBP algorithm in \eqnref{eqn:ccbp},
    \item \textbf{PoIBOMP} - BOMP with polar interpolation, and
    \item \textbf{PaIBOMP+CCBP} - BOMP with parabolic interpolation, where the
        estimates are refined using the CCBP algorithm.
\end{itemize}
We investigate the effect of the optional CCBP
algorithm after the greedy algorithm. To distinguish between the IBOMP algorithm
with and without this optional step, we write IBOMP+CCBP if the algorithm uses
the CCBP algorithm and IBOMP if it does not. Furthermore, to distinguish
whether parabolic interpolation or polar interpolation is used, we use PaIBOMP
and PoIBOMP instead of IBOMP. 
The reason why we use parabolic interpolation in the PaIBOMP+CCBP algorithm,
instead of polar interpolation is that the parabolic interpolation is more
stable when the pulses are overlapping. This is shown in the numerical
experiments.  For the PaIBOMP+CCBP algorithm we set $\xi=0$, as we have rarely
observed that the greedy algorithms chooses the wrong atom for the TDE problem. We have not included CBP~\cite{Ekanadham2011} in our comparison because it is not able to work with complex or negative amplitude coefficients. While the BISP algorithm proposed in~\cite{Fyhn2013} uses CBP, we have modified it by using CCBP so that it can be applied to complex signals.

We use the CVX package \cite{cvx} for solving the convex optimization problems. CVX is a general convex optimization problem solver, which is very good for prototyping; however, if a more customized solver were employed, we believe the computational cost is likely to be decreased.

\subsubsection*{Case A: Well-spaced pulses}
\begin{figure*}
        \centering
        \setlength\figureheight{4.5cm}
        \setlength\figurewidth{0.35\textwidth}
        \includegraphics[width=0.9\textwidth]{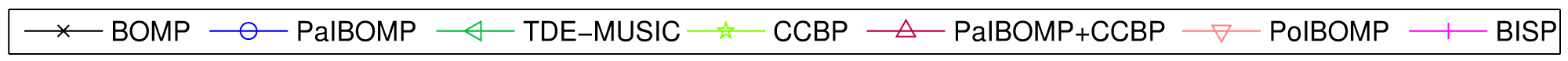}\\
      	\subfloat{
%
%
%

\definecolor{mycolor1}{rgb}{0,0.75,0.25}
\definecolor{mycolor2}{rgb}{0.5,1,0}
\definecolor{mycolor3}{rgb}{1,0.5,0.5}
\definecolor{mycolor4}{rgb}{1,0,1}

\begin{tikzpicture}

\begin{axis}[%
width=\figurewidth,
height=\figureheight,
scale only axis,
xmin=0.05,
xmax=0.5,
xlabel={$\kappa$},
ymode=log,
ymin=0.001,
ymax=1000,
yminorticks=true,
ylabel={Function norm-2 error}
]
\addplot [
color=black,
solid,
mark=x,
mark options={solid},
forget plot
]
table[row sep=crcr]{
0.05 210.879150778922\\
0.1 135.135198764839\\
0.15 54.9945932496728\\
0.2 28.074444111042\\
0.25 11.9010945124078\\
0.3 11.5612482932102\\
0.35 9.59069981244798\\
0.4 9.64823217626994\\
0.45 8.71765961365951\\
0.5 10.5893146367562\\
};
\addplot [
color=blue,
solid,
mark=o,
mark options={solid},
forget plot
]
table[row sep=crcr]{
0.05 210.02066191684\\
0.1 130.975078385858\\
0.15 48.6258208255869\\
0.2 20.0677282368098\\
0.25 3.07180015369541\\
0.3 2.50038612570691\\
0.35 0.589711085677216\\
0.4 0.451296239817775\\
0.45 0.23846073514559\\
0.5 0.228081355853033\\
};
\addplot [
color=mycolor1,
solid,
mark=triangle,
mark options={solid,,rotate=90},
forget plot
]
table[row sep=crcr]{
0.05 167.514255063839\\
0.1 35.7677065395414\\
0.15 3.08006246519731\\
0.2 1.5575926027409\\
0.25 1.10044059641383\\
0.3 0.810859535734223\\
0.35 0.543292434210259\\
0.4 0.386456688469402\\
0.45 0.221873085299581\\
0.5 0.17553101693969\\
};
\addplot [
color=mycolor2,
solid,
mark=star,
mark options={solid},
forget plot
]
table[row sep=crcr]{
0.05 170.102476029866\\
0.1 41.2695027531585\\
0.15 1.24549604039766\\
0.2 0.0215776408747067\\
0.25 0.0125165051167387\\
0.3 0.00890973535207621\\
0.35 0.00626679163688359\\
0.4 0.0053243247008568\\
0.45 0.00444014401283337\\
0.5 0.00399659922244724\\
};
\addplot [
color=purple,
solid,
mark=triangle,
mark options={solid},
forget plot
]
table[row sep=crcr]{
0.05 243.125846991053\\
0.1 134.600957469825\\
0.15 47.8856155799733\\
0.2 19.3193748335676\\
0.25 1.90088976694825\\
0.3 1.76566195722239\\
0.35 0.0076959760092056\\
0.4 0.0078065286848426\\
0.45 0.00581070113300847\\
0.5 0.0058922923318932\\
};
\addplot [
color=mycolor3,
solid,
mark=triangle,
mark options={solid,,rotate=180},
forget plot
]
table[row sep=crcr]{
0.05 216.435398083622\\
0.1 119.715713633582\\
0.15 43.9719596461721\\
0.2 17.9614653939925\\
0.25 1.802763176159\\
0.3 1.66088650011118\\
0.35 0.00278002577517619\\
0.4 0.00275078866208824\\
0.45 0.00250685398086943\\
0.5 0.00259276020254096\\
};
\addplot [
color=mycolor4,
solid,
mark=+,
mark options={solid},
forget plot
]
table[row sep=crcr]{
0.05 284.527802642033\\
0.1 154.352580394839\\
0.15 50.9450354293979\\
0.2 19.5744842862216\\
0.25 1.92633358859402\\
0.3 1.84909205169297\\
0.35 0.00657014184530118\\
0.4 0.00587311888501454\\
0.45 0.00510169669549226\\
0.5 0.00487417325544961\\
};
\end{axis}
\end{tikzpicture}%
}
      	\subfloat{
%
%
%

\definecolor{mycolor1}{rgb}{0,0.75,0.25}
\definecolor{mycolor2}{rgb}{0.5,1,0}
\definecolor{mycolor3}{rgb}{1,0.5,0.5}
\definecolor{mycolor4}{rgb}{1,0,1}

\begin{tikzpicture}

\begin{axis}[%
width=\figurewidth,
height=\figureheight,
scale only axis,
xmin=0,
xmax=50,
xlabel={SNR [dB]},
ymode=log,
ymin=0.001,
ymax=1000,
yminorticks=true,
ylabel={Function norm-2 error}
]
\addplot [
color=black,
solid,
mark=x,
mark options={solid},
forget plot
]
table[row sep=crcr]{
0 18.9131114113809\\
5 12.6338788050632\\
10 9.69010895990208\\
15 9.19475766703271\\
20 9.28348828114053\\
25 8.92391106287804\\
30 10.7781872181921\\
35 9.14400364238099\\
40 11.2200623865213\\
45 9.14453963134405\\
50 9.36074154457451\\
};
\addplot [
color=blue,
solid,
mark=o,
mark options={solid},
forget plot
]
table[row sep=crcr]{
0 9.00276980609965\\
5 2.77626321174672\\
10 0.898519526805411\\
15 0.568116730139921\\
20 0.453800477569446\\
25 0.517867816425928\\
30 0.403191215934593\\
35 0.387563281356888\\
40 0.481666680341415\\
45 0.470350464759203\\
50 0.517632476907017\\
};
\addplot [
color=mycolor1,
solid,
mark=triangle,
mark options={solid,,rotate=90},
forget plot
]
table[row sep=crcr]{
0 214.000699704458\\
5 30.0989587820965\\
10 4.19503498568923\\
15 1.50123689585252\\
20 0.853388830406641\\
25 0.576639455345901\\
30 0.497195587089801\\
35 0.34697624425545\\
40 0.292948192262091\\
45 0.322257773327781\\
50 0.315650243572654\\
};
\addplot [
color=mycolor2,
solid,
mark=star,
mark options={solid},
forget plot
]
table[row sep=crcr]{
0 29.7178784558787\\
5 11.928537890621\\
10 3.93044911936603\\
15 1.39309566428871\\
20 0.496316284518743\\
25 0.286486481624524\\
30 0.0626289251625042\\
35 0.02692096308786\\
40 0.0154341295798334\\
45 0.0101520858831258\\
50 0.00812394525193096\\
};
\addplot [
color=purple,
solid,
mark=triangle,
mark options={solid},
forget plot
]
table[row sep=crcr]{
0 28.0512682620892\\
5 11.5658537062527\\
10 3.7006903255875\\
15 1.352785083244\\
20 0.489093055944433\\
25 0.283686089474075\\
30 0.0653348846954051\\
35 0.0289778626104986\\
40 0.0199848061398684\\
45 0.0121923681179446\\
50 0.00886478997054369\\
};
\addplot [
color=mycolor3,
solid,
mark=triangle,
mark options={solid,,rotate=180},
forget plot
]
table[row sep=crcr]{
0 11.2054598218977\\
5 2.98289329424053\\
10 0.692271221398525\\
15 0.254964145829397\\
20 0.0816747364255563\\
25 0.0249956968498206\\
30 0.010705985597816\\
35 0.00515048923375024\\
40 0.00396588383118157\\
45 0.00276420340040088\\
50 0.00279723834397953\\
};
\addplot [
color=mycolor4,
solid,
mark=+,
mark options={solid},
forget plot
]
table[row sep=crcr]{
0 27.9574433282565\\
5 11.6566253314122\\
10 3.74024268432088\\
15 1.35411010096986\\
20 0.483336788087259\\
25 0.282381115390577\\
30 0.0614695490770871\\
35 0.0265591858852138\\
40 0.0154781393351408\\
45 0.0101991642148171\\
50 0.00832495966708263\\
};
\end{axis}
\end{tikzpicture}%
}\\
      	\subfloat{
%
%
%

\definecolor{mycolor1}{rgb}{0,0.75,0.25}
\definecolor{mycolor2}{rgb}{0.5,1,0}
\definecolor{mycolor3}{rgb}{1,0.5,0.5}
\definecolor{mycolor4}{rgb}{1,0,1}

\begin{tikzpicture}

\begin{axis}[%
width=\figurewidth,
height=\figureheight,
scale only axis,
xmin=0.05,
xmax=0.5,
xlabel={$\kappa$},
ymode=log,
ymin=1e-09,
ymax=10,
yminorticks=true,
ylabel={$\text{\vb-MSE [}\mu{}\text{s}^\text{2}\text{]}$}
]
\addplot [
color=black,
solid,
mark=x,
mark options={solid},
forget plot
]
table[row sep=crcr]{
0.05 1.24405573578914\\
0.1 0.123785750674304\\
0.15 0.00565214931069251\\
0.2 0.00293289161367181\\
0.25 0.000555442882126295\\
0.3 0.000180107234091337\\
0.35 3.46877776208205e-05\\
0.4 3.39104261084477e-05\\
0.45 3.30052300134785e-05\\
0.5 3.55653413583309e-05\\
};
\addplot [
color=blue,
solid,
mark=o,
mark options={solid},
forget plot
]
table[row sep=crcr]{
0.05 1.24191768888227\\
0.1 0.122696635055072\\
0.15 0.00563313795388682\\
0.2 0.00292781478081207\\
0.25 0.000531450766700273\\
0.3 0.000144428366215928\\
0.35 1.82289126255364e-06\\
0.4 1.53032517543294e-06\\
0.45 8.4298356380768e-07\\
0.5 7.63464429407491e-07\\
};
\addplot [
color=mycolor1,
solid,
mark=triangle,
mark options={solid,,rotate=90},
forget plot
]
table[row sep=crcr]{
0.05 1.15134647205771\\
0.1 0.00249933678550569\\
0.15 7.04206207407864e-05\\
0.2 5.83281948028958e-06\\
0.25 6.1058943679765e-06\\
0.3 6.6304915390512e-06\\
0.35 6.17293302046594e-06\\
0.4 6.06503159254718e-06\\
0.45 6.06225191315026e-06\\
0.5 5.85062214166331e-06\\
};
\addplot [
color=mycolor2,
solid,
mark=star,
mark options={solid},
forget plot
]
table[row sep=crcr]{
0.05 1.79455416904502\\
0.1 0.170706237788758\\
0.15 6.7354958324678e-05\\
0.2 5.12914025099031e-08\\
0.25 3.31475959778146e-08\\
0.3 2.160264026462e-08\\
0.35 1.6567009929232e-08\\
0.4 1.59090140269109e-08\\
0.45 1.68315126127257e-08\\
0.5 1.40364619666587e-08\\
};
\addplot [
color=purple,
solid,
mark=triangle,
mark options={solid},
forget plot
]
table[row sep=crcr]{
0.05 1.2443827718909\\
0.1 0.12371187644528\\
0.15 0.00534717629848064\\
0.2 0.00293113347864992\\
0.25 0.000536427572130348\\
0.3 0.000171754259622415\\
0.35 1.62561343259895e-08\\
0.4 1.69342535818029e-08\\
0.45 1.49658699792031e-08\\
0.5 1.12766945570602e-08\\
};
\addplot [
color=mycolor3,
solid,
mark=triangle,
mark options={solid,,rotate=180},
forget plot
]
table[row sep=crcr]{
0.05 1.24151239699607\\
0.1 0.120747970969693\\
0.15 0.00475677690789837\\
0.2 0.0025387005636941\\
0.25 0.000436481374037436\\
0.3 0.000124566018270078\\
0.35 9.14383199723028e-09\\
0.4 9.34919589989084e-09\\
0.45 8.82447773829675e-09\\
0.5 8.57547465851134e-09\\
};
\addplot [
color=mycolor4,
solid,
mark=+,
mark options={solid},
forget plot
]
table[row sep=crcr]{
0.05 1.99769654118041\\
0.1 0.512611819042406\\
0.15 0.031573865366597\\
0.2 0.00302147790497996\\
0.25 0.000550248618793244\\
0.3 0.000180626622940029\\
0.35 1.53702752590487e-08\\
0.4 1.34353126755273e-08\\
0.45 1.40424708330461e-08\\
0.5 1.07242472678377e-08\\
};
\end{axis}
\end{tikzpicture}%
}
      	\subfloat{
%
%
%

\definecolor{mycolor1}{rgb}{0,0.75,0.25}
\definecolor{mycolor2}{rgb}{0.5,1,0}
\definecolor{mycolor3}{rgb}{1,0.5,0.5}
\definecolor{mycolor4}{rgb}{1,0,1}

\begin{tikzpicture}

\begin{axis}[%
width=\figurewidth,
height=\figureheight,
scale only axis,
xmin=0,
xmax=50,
xlabel={SNR [dB]},
ymode=log,
ymin=1e-09,
ymax=10,
yminorticks=true,
ylabel={$\text{\vb-MSE [}\mu{}\text{s}^\text{2}\text{]}$}
]
\addplot [
color=black,
solid,
mark=x,
mark options={solid},
forget plot
]
table[row sep=crcr]{
0 0.286734168813138\\
5 0.000700401061734659\\
10 3.28973146731231e-05\\
15 3.3982104739843e-05\\
20 3.52428614992613e-05\\
25 3.21171359774599e-05\\
30 3.74730325775607e-05\\
35 3.27081958433179e-05\\
40 3.87126882998145e-05\\
45 3.34311633035177e-05\\
50 3.32613637794571e-05\\
};
\addplot [
color=blue,
solid,
mark=o,
mark options={solid},
forget plot
]
table[row sep=crcr]{
0 0.287666953643572\\
5 0.000655928983305622\\
10 2.1685485477038e-06\\
15 1.80481738754778e-06\\
20 1.42704637726801e-06\\
25 1.70401982380304e-06\\
30 1.49694473781618e-06\\
35 1.22227570817594e-06\\
40 1.42225746272692e-06\\
45 1.53638175645889e-06\\
50 1.63333732366022e-06\\
};
\addplot [
color=mycolor1,
solid,
mark=triangle,
mark options={solid,,rotate=90},
forget plot
]
table[row sep=crcr]{
0 2.85250180284459\\
5 0.349333472614424\\
10 8.62749027646609e-06\\
15 6.78526723491635e-06\\
20 5.98286721056646e-06\\
25 6.43875324605677e-06\\
30 6.30532981266445e-06\\
35 5.9679426707855e-06\\
40 6.17871462451112e-06\\
45 6.09918754405734e-06\\
50 6.1423577162979e-06\\
};
\addplot [
color=mycolor2,
solid,
mark=star,
mark options={solid},
forget plot
]
table[row sep=crcr]{
0 0.472444475923572\\
5 0.0400905989580168\\
10 6.62273406965622e-06\\
15 4.22106074879324e-06\\
20 1.4002697956201e-06\\
25 7.29730319215511e-07\\
30 1.62317355358474e-07\\
35 5.24637783556942e-08\\
40 3.29711244347641e-08\\
45 2.6537817186568e-08\\
50 2.26187228563017e-08\\
};
\addplot [
color=purple,
solid,
mark=triangle,
mark options={solid},
forget plot
]
table[row sep=crcr]{
0 0.288680998152658\\
5 0.00066172929125183\\
10 5.40846273616696e-06\\
15 2.53450429614991e-06\\
20 1.29793423657865e-06\\
25 6.93562283599538e-07\\
30 1.64006980038168e-07\\
35 5.71274169763625e-08\\
40 3.18176434074307e-08\\
45 3.18484131739725e-08\\
50 2.19387279997506e-08\\
};
\addplot [
color=mycolor3,
solid,
mark=triangle,
mark options={solid,,rotate=180},
forget plot
]
table[row sep=crcr]{
0 0.287074213344005\\
5 0.000607485381342383\\
10 2.31701740984004e-06\\
15 8.9288432116165e-07\\
20 2.73236400718434e-07\\
25 6.62295661899118e-08\\
30 3.27705753075001e-08\\
35 1.62109294894225e-08\\
40 1.23452618790623e-08\\
45 9.46763247797294e-09\\
50 9.8760329579839e-09\\
};
\addplot [
color=mycolor4,
solid,
mark=+,
mark options={solid},
forget plot
]
table[row sep=crcr]{
0 0.239686667314417\\
5 0.00065877322303761\\
10 6.57605842568377e-06\\
15 3.44016080480813e-06\\
20 1.39964141762286e-06\\
25 7.22960779062097e-07\\
30 1.61193416722215e-07\\
35 5.09886464151618e-08\\
40 4.8447370705817e-08\\
45 2.64509807022369e-08\\
50 2.1842677007666e-08\\
};
\end{axis}
\end{tikzpicture}%
}\\
      	\subfloat{
%
%
%

\definecolor{mycolor1}{rgb}{0,0.75,0.25}
\definecolor{mycolor2}{rgb}{0.5,1,0}
\definecolor{mycolor3}{rgb}{1,0.5,0.5}
\definecolor{mycolor4}{rgb}{1,0,1}

\begin{tikzpicture}

\begin{axis}[%
width=\figurewidth,
height=\figureheight,
scale only axis,
xmin=0.05,
xmax=0.5,
xlabel={$\kappa$},
ymode=log,
ymin=0.01,
ymax=10000,
yminorticks=true,
ylabel={Average computation time [s]}
]
\addplot [
color=black,
solid,
mark=x,
mark options={solid},
forget plot
]
table[row sep=crcr]{
0.05 0.0338428402777778\\
0.1 0.0390921458333333\\
0.15 0.0481826527777778\\
0.2 0.0597886111111111\\
0.25 0.0674526388888889\\
0.3 0.0721861597222222\\
0.35 0.0775155208333334\\
0.4 0.0906920347222221\\
0.45 0.0891849166666666\\
0.5 0.0988827291666666\\
};
\addplot [
color=blue,
solid,
mark=o,
mark options={solid},
forget plot
]
table[row sep=crcr]{
0.05 0.0269427291666667\\
0.1 0.0379374583333333\\
0.15 0.0481153819444445\\
0.2 0.0572444652777778\\
0.25 0.0681738194444445\\
0.3 0.0731846666666667\\
0.35 0.0758213402777778\\
0.4 0.0856764305555555\\
0.45 0.0914088888888889\\
0.5 0.097047875\\
};
\addplot [
color=mycolor1,
solid,
mark=triangle,
mark options={solid,,rotate=90},
forget plot
]
table[row sep=crcr]{
0.05 2.04704118055556\\
0.1 2.08834422916667\\
0.15 3.01234143055556\\
0.2 3.3431075\\
0.25 3.64957774305555\\
0.3 4.21163670833333\\
0.35 4.78119986111111\\
0.4 5.36285274305556\\
0.45 6.05788901388889\\
0.5 6.71637430555556\\
};
\addplot [
color=mycolor2,
solid,
mark=star,
mark options={solid},
forget plot
]
table[row sep=crcr]{
0.05 852.344900965278\\
0.1 877.558536055556\\
0.15 905.9254783125\\
0.2 936.561225083333\\
0.25 971.057311006945\\
0.3 1015.08690521528\\
0.35 1037.97411115278\\
0.4 1071.62520424306\\
0.45 1014.62230402778\\
0.5 1027.56356902083\\
};
\addplot [
color=purple,
solid,
mark=triangle,
mark options={solid},
forget plot
]
table[row sep=crcr]{
0.05 4.30339548611111\\
0.1 4.78099402777778\\
0.15 4.95861685416667\\
0.2 4.96115043055555\\
0.25 5.15543459722222\\
0.3 5.04127640277778\\
0.35 5.11912548611111\\
0.4 5.29771659027778\\
0.45 5.21882213194444\\
0.5 5.24319228472222\\
};
\addplot [
color=mycolor3,
solid,
mark=triangle,
mark options={solid,,rotate=180},
forget plot
]
table[row sep=crcr]{
0.05 0.0455035069444444\\
0.1 0.0600158472222222\\
0.15 0.0699078819444444\\
0.2 0.0816292638888889\\
0.25 0.0951343541666667\\
0.3 0.0949144930555555\\
0.35 0.101246402777778\\
0.4 0.124234430555556\\
0.45 0.1312314375\\
0.5 0.138058368055556\\
};
\addplot [
color=mycolor4,
solid,
mark=+,
mark options={solid},
forget plot
]
table[row sep=crcr]{
0.05 10.8381452575486\\
0.1 10.6793531549444\\
0.15 11.0101945821111\\
0.2 11.1182505762639\\
0.25 11.3447861499861\\
0.3 11.4692950523611\\
0.35 11.5604363813055\\
0.4 11.677586605\\
0.45 11.7669349165625\\
0.5 11.7906384479236\\
};
\end{axis}
\end{tikzpicture}%
}
      	\subfloat{
%
%
%

\definecolor{mycolor1}{rgb}{0,0.75,0.25}
\definecolor{mycolor2}{rgb}{0.5,1,0}
\definecolor{mycolor3}{rgb}{1,0.5,0.5}
\definecolor{mycolor4}{rgb}{1,0,1}

\begin{tikzpicture}

\begin{axis}[%
width=\figurewidth,
height=\figureheight,
scale only axis,
xmin=0,
xmax=50,
xlabel={SNR [dB]},
ymode=log,
ymin=0.01,
ymax=10000,
yminorticks=true,
ylabel={Average computation time [s]}
]
\addplot [
color=black,
solid,
mark=x,
mark options={solid},
forget plot
]
table[row sep=crcr]{
0 0.108864521008403\\
5 0.0851463277310924\\
10 0.0877740504201681\\
15 0.0868106302521008\\
20 0.0840229159663866\\
25 0.0822973781512605\\
30 0.087889949579832\\
35 0.083393974789916\\
40 0.0843166134453782\\
45 0.0848015210084034\\
50 0.083806756302521\\
};
\addplot [
color=blue,
solid,
mark=o,
mark options={solid},
forget plot
]
table[row sep=crcr]{
0 0.0964360504201681\\
5 0.0905745546218487\\
10 0.0908752352941176\\
15 0.0882136050420168\\
20 0.0832881848739496\\
25 0.0898507647058824\\
30 0.0855819243697479\\
35 0.0859053025210084\\
40 0.0818638487394958\\
45 0.0827334285714286\\
50 0.0816109159663865\\
};
\addplot [
color=mycolor1,
solid,
mark=triangle,
mark options={solid,,rotate=90},
forget plot
]
table[row sep=crcr]{
0 5.58063643697479\\
5 3.92428306722689\\
10 4.40661461344538\\
15 4.7131102184874\\
20 4.7390023697479\\
25 4.84997802521008\\
30 4.93415905042017\\
35 5.17750602521008\\
40 5.15029211764706\\
45 5.21740531932773\\
50 5.29709160504202\\
};
\addplot [
color=mycolor2,
solid,
mark=star,
mark options={solid},
forget plot
]
table[row sep=crcr]{
0 979.402827840336\\
5 997.530847840336\\
10 1020.05179169748\\
15 1037.28650609244\\
20 1051.49264117647\\
25 1050.28769759664\\
30 1060.19971432773\\
35 1051.48097878992\\
40 1053.89036869748\\
45 1052.82326763866\\
50 1060.31160711765\\
};
\addplot [
color=purple,
solid,
mark=triangle,
mark options={solid},
forget plot
]
table[row sep=crcr]{
0 4.51037710084034\\
5 4.49420488235294\\
10 4.65409842016807\\
15 4.79055778991597\\
20 5.04070487394958\\
25 4.94710473109244\\
30 5.07602546218487\\
35 4.97389149579832\\
40 5.0142753697479\\
45 5.10920889915967\\
50 5.24341168907563\\
};
\addplot [
color=mycolor3,
solid,
mark=triangle,
mark options={solid,,rotate=180},
forget plot
]
table[row sep=crcr]{
0 0.145393344537815\\
5 0.144097344537815\\
10 0.131816613445378\\
15 0.135616596638655\\
20 0.128198361344538\\
25 0.127823243697479\\
30 0.130102386554622\\
35 0.132410571428571\\
40 0.121773285714286\\
45 0.130089579831933\\
50 0.130560865546219\\
};
\addplot [
color=mycolor4,
solid,
mark=+,
mark options={solid},
forget plot
]
table[row sep=crcr]{
0 5.3760574360084\\
5 8.15136464951261\\
10 10.1927801932017\\
15 11.570718193479\\
20 11.8162977626807\\
25 11.8329421717227\\
30 11.8272263787815\\
35 11.7500904514454\\
40 11.7363821578151\\
45 11.7802900059244\\
50 11.8154562607227\\
};
\end{axis}
\end{tikzpicture}%
}
        \caption{The estimator precision with non-overlapping pulses for the TDE
            problem. The left figures are noiseless experiments for varying
            choices of subsampling ratios, $\kappa$, while the right figures are
            for $\kappa=0.4$ and with varying SNR levels. The top figures is
            signal reconstruction quality, the middle row is translation
            parameter estimation precision and the bottom row is average 
            computation time.}
        \label{fig:tde_K3_nonoverlapping_pulses}
\end{figure*}

This experiment is performed with $K=3$ well-spaced pulses, i.e., $\eta=0$.
The minimum separation between pulses is set to $10^{-6}$ seconds, i.e., exactly
the width of a pulse. This means there is no overlap anywhere between pulses.
The result of our comparison is shown in
\figref{fig:tde_K3_nonoverlapping_pulses}.
As shown, the polar interpolation algorithms outperform all the other
algorithms.  With $\xi=0$ PaIBOMP+CCBP has the same computational complexity as
TDE-MUSIC, whereas CCBP is significantly more computationally heavy.

Also note that PoIBOMP outperforms BISP, CCBP and PaIBOMP+CCBP
while also being significantly less computational complex. This is because the
pulses are well separated. In the next experiment, we use overlapping pulses
which affects the purely greedy algorithms more than the pure and hybrid convex
optimization algorithms.

\subsubsection*{Case B: Overlapping pulses}
\begin{figure*}
        \centering
        \setlength\figureheight{4.5cm}
        \setlength\figurewidth{0.35\textwidth}
        \includegraphics[width=0.9\textwidth]{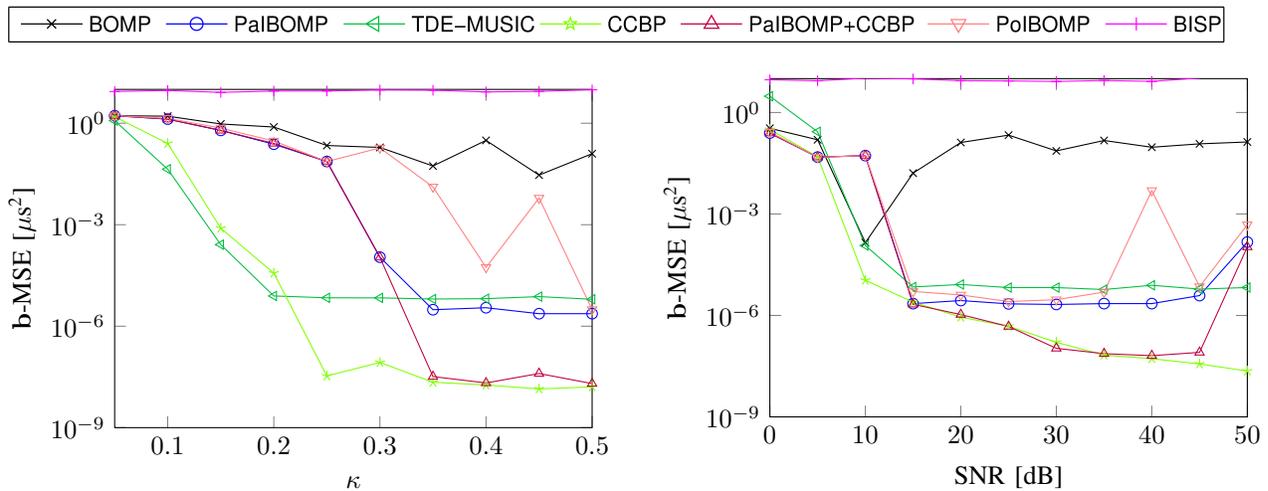}\\
      	\subfloat{
%
%
%

\definecolor{mycolor1}{rgb}{0,0.75,0.25}
\definecolor{mycolor2}{rgb}{0.5,1,0}
\definecolor{mycolor3}{rgb}{1,0.5,0.5}
\definecolor{mycolor4}{rgb}{1,0,1}

\begin{tikzpicture}

\begin{axis}[%
width=\figurewidth,
height=\figureheight,
scale only axis,
xmin=0.05,
xmax=0.5,
xlabel={$\kappa$},
ymode=log,
ymin=1e-09,
ymax=10,
yminorticks=true,
ylabel={$\text{\vb-MSE [}\mu{}\text{s}^\text{2}\text{]}$}
]
\addplot [
color=black,
solid,
mark=x,
mark options={solid},
forget plot
]
table[row sep=crcr]{
0.05 1.66044595499238\\
0.1 1.61721816734328\\
0.15 0.951493495691867\\
0.2 0.771626994620259\\
0.25 0.219681476618333\\
0.3 0.19271832143602\\
0.35 0.0544434978724581\\
0.4 0.31091287446288\\
0.45 0.0293372176246042\\
0.5 0.125637381624593\\
};
\addplot [
color=blue,
solid,
mark=o,
mark options={solid},
forget plot
]
table[row sep=crcr]{
0.05 1.66709545277092\\
0.1 1.32425984812692\\
0.15 0.613818802793945\\
0.2 0.240028841525699\\
0.25 0.0739033546627265\\
0.3 0.000109651888430372\\
0.35 3.0691183793411e-06\\
0.4 3.51196358000139e-06\\
0.45 2.3404411771635e-06\\
0.5 2.34213096538619e-06\\
};
\addplot [
color=mycolor1,
solid,
mark=triangle,
mark options={solid,,rotate=90},
forget plot
]
table[row sep=crcr]{
0.05 1.19468423312907\\
0.1 0.0437008511053529\\
0.15 0.000257550155270827\\
0.2 7.82490254285942e-06\\
0.25 6.89731374624059e-06\\
0.3 6.86071178107458e-06\\
0.35 6.35865053621188e-06\\
0.4 6.50813912691128e-06\\
0.45 7.49019704682557e-06\\
0.5 6.24027189983621e-06\\
};
\addplot [
color=mycolor2,
solid,
mark=star,
mark options={solid},
forget plot
]
table[row sep=crcr]{
0.05 1.55582402188597\\
0.1 0.252704650281543\\
0.15 0.00078619341166916\\
0.2 3.71032104445198e-05\\
0.25 3.37202228687325e-08\\
0.3 8.46521810110583e-08\\
0.35 2.2023331742258e-08\\
0.4 1.82893789039679e-08\\
0.45 1.39532874561235e-08\\
0.5 1.61123741524884e-08\\
};
\addplot [
color=purple,
solid,
mark=triangle,
mark options={solid},
forget plot
]
table[row sep=crcr]{
0.05 1.66761830956022\\
0.1 1.32393302986831\\
0.15 0.613754654461764\\
0.2 0.253579194415144\\
0.25 0.073783251601522\\
0.3 0.00010850515573056\\
0.35 3.2363211276326e-08\\
0.4 2.09850734187423e-08\\
0.45 3.98420254329237e-08\\
0.5 2.00875797109645e-08\\
};
\addplot [
color=mycolor3,
solid,
mark=triangle,
mark options={solid,,rotate=180},
forget plot
]
table[row sep=crcr]{
0.05 1.67609303085968\\
0.1 1.39633237504843\\
0.15 0.717709681872388\\
0.2 0.299201212624633\\
0.25 0.0737885534786604\\
0.3 0.18643130873334\\
0.35 0.0131051668633472\\
0.4 5.46769473229494e-05\\
0.45 0.00614476059332909\\
0.5 3.10559119088645e-06\\
};
\addplot [
color=mycolor4,
solid,
mark=+,
mark options={solid},
forget plot
]
table[row sep=crcr]{
0.05 8.68885258282837\\
0.1 9.27037044458446\\
0.15 8.17328075689216\\
0.2 8.94488214191817\\
0.25 8.98075198717181\\
0.3 9.60257431778716\\
0.35 9.44411301383489\\
0.4 8.42246255746595\\
0.45 8.75410155215743\\
0.5 9.73341483607586\\
};
\end{axis}
\end{tikzpicture}%
}
      	\subfloat{
%
%
%

\definecolor{mycolor1}{rgb}{0,0.75,0.25}
\definecolor{mycolor2}{rgb}{0.5,1,0}
\definecolor{mycolor3}{rgb}{1,0.5,0.5}
\definecolor{mycolor4}{rgb}{1,0,1}

\begin{tikzpicture}

\begin{axis}[%
width=\figurewidth,
height=\figureheight,
scale only axis,
xmin=0,
xmax=50,
xlabel={SNR [dB]},
ymode=log,
ymin=1e-09,
ymax=10,
yminorticks=true,
ylabel={$\text{\vb-MSE [}\mu{}\text{s}^\text{2}\text{]}$}
]
\addplot [
color=black,
solid,
mark=x,
mark options={solid},
forget plot
]
table[row sep=crcr]{
0 0.340690728861053\\
5 0.156141531408622\\
10 0.000141736083192517\\
15 0.0161104224085839\\
20 0.12962782008209\\
25 0.213839780994396\\
30 0.0730757386570873\\
35 0.14848335005896\\
40 0.0938179156984822\\
45 0.118166092717302\\
50 0.134257714346441\\
};
\addplot [
color=blue,
solid,
mark=o,
mark options={solid},
forget plot
]
table[row sep=crcr]{
0 0.245701393375455\\
5 0.0478776725428606\\
10 0.0529872033443249\\
15 2.23265962772014e-06\\
20 2.75255906149867e-06\\
25 2.19601262954139e-06\\
30 2.10984038409227e-06\\
35 2.23321625056095e-06\\
40 2.23402195206301e-06\\
45 3.85761439345297e-06\\
50 0.000147572195386545\\
};
\addplot [
color=mycolor1,
solid,
mark=triangle,
mark options={solid,,rotate=90},
forget plot
]
table[row sep=crcr]{
0 3.02806258807922\\
5 0.269458836562811\\
10 0.000116879524302759\\
15 6.93186052039542e-06\\
20 8.22562589611755e-06\\
25 6.64189417755524e-06\\
30 6.64019269831113e-06\\
35 5.82618637840124e-06\\
40 7.78028098550184e-06\\
45 5.99467220420166e-06\\
50 6.67919603001012e-06\\
};
\addplot [
color=mycolor2,
solid,
mark=star,
mark options={solid},
forget plot
]
table[row sep=crcr]{
0 0.313703411334536\\
5 0.0488580031222314\\
10 1.09493087009709e-05\\
15 2.4766847428473e-06\\
20 8.87421235237156e-07\\
25 4.77069087853786e-07\\
30 1.60215415237798e-07\\
35 6.71716705709684e-08\\
40 5.25549466064243e-08\\
45 3.67611643135636e-08\\
50 2.26658054837257e-08\\
};
\addplot [
color=purple,
solid,
mark=triangle,
mark options={solid},
forget plot
]
table[row sep=crcr]{
0 0.245826995234414\\
5 0.0479041998572848\\
10 0.0527856200172233\\
15 2.13068252747997e-06\\
20 1.06117965454064e-06\\
25 4.70634059830757e-07\\
30 1.06571981843045e-07\\
35 7.41599577732926e-08\\
40 6.45275899631284e-08\\
45 8.08783096680943e-08\\
50 0.000106180528038392\\
};
\addplot [
color=mycolor3,
solid,
mark=triangle,
mark options={solid,,rotate=180},
forget plot
]
table[row sep=crcr]{
0 0.263158855069458\\
5 0.0474348617216764\\
10 0.0531171968528866\\
15 5.09775805629659e-06\\
20 4.02792967453458e-06\\
25 2.57191373809061e-06\\
30 2.91253564389763e-06\\
35 4.88698038483551e-06\\
40 0.00492861902119217\\
45 7.11383439028309e-06\\
50 0.000478886570045936\\
};
\addplot [
color=mycolor4,
solid,
mark=+,
mark options={solid},
forget plot
]
table[row sep=crcr]{
0 9.28206842859513\\
5 8.65387611286972\\
10 10.2978236000464\\
15 9.83265205931446\\
20 8.7256377891271\\
25 8.56157531074144\\
30 8.2712067941656\\
35 8.8178834161572\\
40 8.31224869900532\\
45 10.4138049974169\\
};
\end{axis}
\end{tikzpicture}%
}
        \vspace{-2mm}
        \caption{Estimator precision with overlapping pulses for the TDE
        problem. Left: noiseless case with varying subsampling ratios;
        right: subsampling ratio $\kappa=0.4$ with varying SNR levels.}
        \label{fig:tde_K3_overlapping_pulses}
        \vspace{-5mm}
\end{figure*}

For this experiment we use the same parameter values, except that the minimum
pulse separation is now set to $5\cdot T_s$, i.e., five times the sampling rate.
The reason why we do not set the separation to $0$ is that if two identical
pulses are received, there is no possibility of correctly decoding these without
introducing further assumptions. Therefore, we introduce this minimum spacing.
We set $\eta=1$, i.e., we disable the band exclusion, such that there is no
restriction on which dictionary atoms are used in each iteration. The result is
shown in \figref{fig:tde_K3_overlapping_pulses}.
As can be seen the greedy algorithm are heavily affected by this, such as
BOMP and PoIBOMP, with BISP having the worst performance due to its incoherence assumption. The PaIBOMP+CCBP algorithm is also affected in that it
requires a slightly higher $\kappa$ before it attains the same performance as
CCBP than in \figref{fig:tde_K3_nonoverlapping_pulses}. This figure also shows
why we use parabolic interpolation in the PaIBOMP+CCBP algorithm, instead of
the polar interpolation from PoIBOMP. When the pulses are overlapping the polar
interpolation has erratic instability issues. It is important to note that the
irregularities for PoIBOMP in \figref{fig:tde_K3_overlapping_pulses} are due to
a single Monte Carlo simulation in which the estimation fails. Polar
interpolation relies on the value of all $N$ dimensions of the signal for the
hypersphere assumption and when pulses are overlapping this assumption is
incorrect. As shown in \eqnref{eqn:poibomp} the translation estimate for
PoIBOMP relies on finding the inverse tangent and if $\hat{x}_2$ is erroneous
this may result in a large error. In contrast, the parabolic interpolation uses
only three points from the cross correlation function. Hence, with overlapping
pulses the PoIBOMP algorithm suffers more from the interference from other
pulses than PaIBOMP.

\subsubsection*{Case C: Noise folding}
\begin{figure*}
        \centering
        \setlength\figureheight{4.5cm}
        \setlength\figurewidth{0.35\textwidth}
        \includegraphics[width=0.9\textwidth]{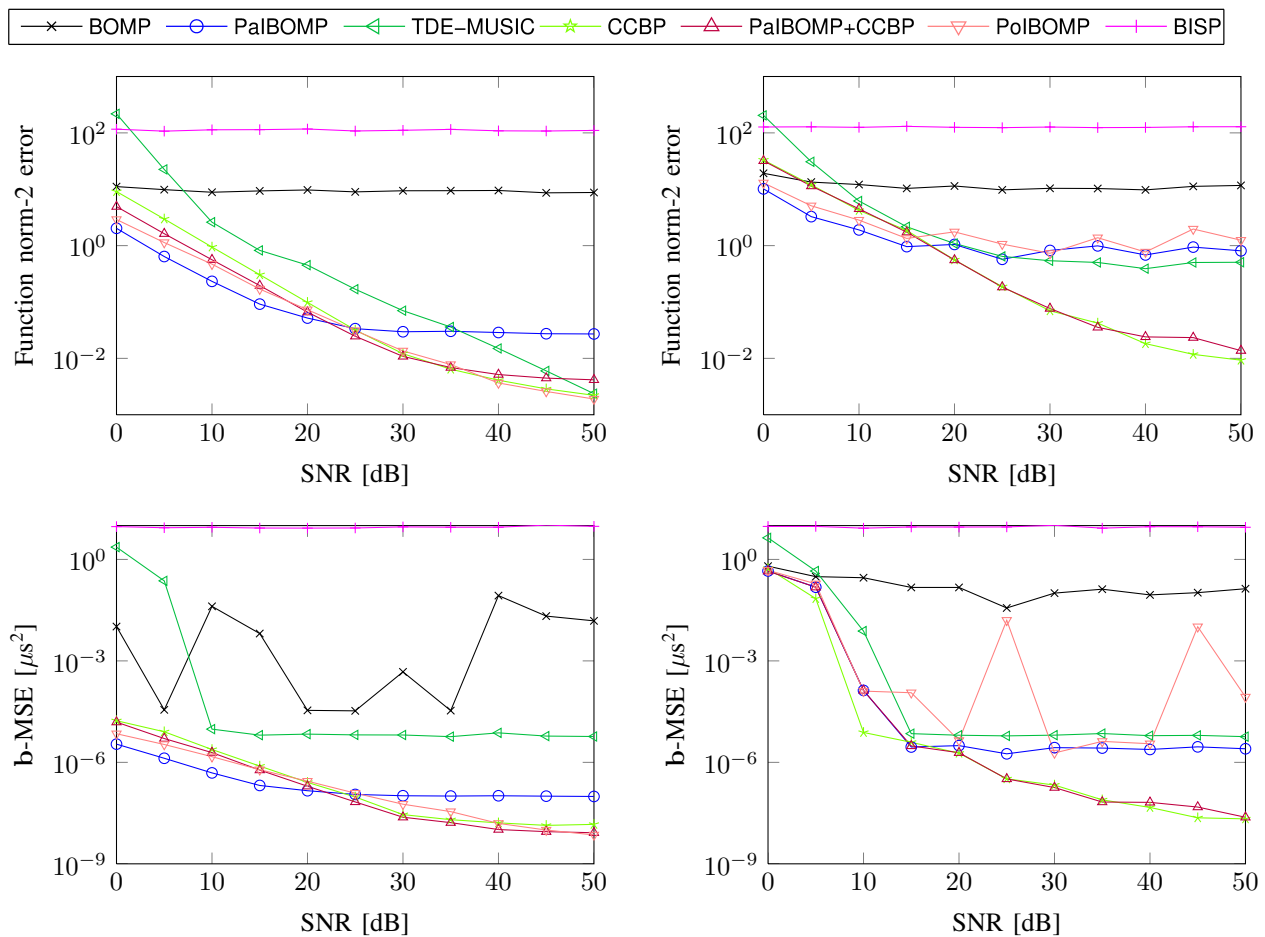}\\
      	\subfloat{
%
%
%

\definecolor{mycolor1}{rgb}{0,0.75,0.25}
\definecolor{mycolor2}{rgb}{0.5,1,0}
\definecolor{mycolor3}{rgb}{1,0.5,0.5}
\definecolor{mycolor4}{rgb}{1,0,1}

\begin{tikzpicture}

\begin{axis}[%
width=\figurewidth,
height=\figureheight,
scale only axis,
xmin=0,
xmax=50,
xlabel={SNR [dB]},
ymode=log,
ymin=0.001,
ymax=1000,
yminorticks=true,
ylabel={Function norm-2 error}
]
\addplot [
color=black,
solid,
mark=x,
mark options={solid},
forget plot
]
table[row sep=crcr]{
0 11.1408326366556\\
5 9.83785810184811\\
10 8.851697421893\\
15 9.34370246512558\\
20 9.71458905726775\\
25 8.98209702022124\\
30 9.40271585328253\\
35 9.43164319553267\\
40 9.51338104925702\\
45 8.63970829494214\\
50 8.75930504928435\\
};
\addplot [
color=blue,
solid,
mark=o,
mark options={solid},
forget plot
]
table[row sep=crcr]{
0 2.02798629940552\\
5 0.640550516677931\\
10 0.232334127143031\\
15 0.0917242446958579\\
20 0.0518656802644209\\
25 0.0336493424567658\\
30 0.0296508765874264\\
35 0.0302310527252409\\
40 0.028749310308105\\
45 0.0273561268321024\\
50 0.0271033290368829\\
};
\addplot [
color=mycolor1,
solid,
mark=triangle,
mark options={solid,,rotate=90},
forget plot
]
table[row sep=crcr]{
0 217.509956631037\\
5 22.4924483516943\\
10 2.60244927231523\\
15 0.823174700477263\\
20 0.450980784039907\\
25 0.168534485501722\\
30 0.0701827944390795\\
35 0.0360009368561416\\
40 0.0149750558592626\\
45 0.00599734436631519\\
50 0.00236341893105853\\
};
\addplot [
color=mycolor2,
solid,
mark=star,
mark options={solid},
forget plot
]
table[row sep=crcr]{
0 9.09927008756072\\
5 2.97058332640092\\
10 0.94339994352499\\
15 0.305150027259545\\
20 0.0983886105536308\\
25 0.0319036095776299\\
30 0.0121002724774337\\
35 0.00642335509465931\\
40 0.00410088172539889\\
45 0.0028771987454732\\
50 0.00221034806628949\\
};
\addplot [
color=purple,
solid,
mark=triangle,
mark options={solid},
forget plot
]
table[row sep=crcr]{
0 4.92648729817161\\
5 1.62803755781042\\
10 0.561246981966776\\
15 0.195586753213807\\
20 0.0660658342670941\\
25 0.0246815312213658\\
30 0.0108367261894103\\
35 0.00687632695937554\\
40 0.00516292359311597\\
45 0.00446436305029989\\
50 0.00416211338001038\\
};
\addplot [
color=mycolor3,
solid,
mark=triangle,
mark options={solid,,rotate=180},
forget plot
]
table[row sep=crcr]{
0 2.90780188643297\\
5 1.1337336476444\\
10 0.469102086591192\\
15 0.169640674673669\\
20 0.0728687673682205\\
25 0.031156237165273\\
30 0.0135576320804488\\
35 0.00776959362050353\\
40 0.00366178270278483\\
45 0.00259053235521718\\
50 0.0019003789422269\\
};
\addplot [
color=mycolor4,
solid,
mark=+,
mark options={solid},
forget plot
]
table[row sep=crcr]{
0 116.217191242788\\
5 107.322560418194\\
10 113.438191283872\\
15 114.110929084719\\
20 117.145456991646\\
25 108.053593917178\\
30 111.182951212588\\
35 115.27188961042\\
40 108.73520175924\\
45 107.914101209599\\
50 110.513805045146\\
};
\end{axis}
\end{tikzpicture}%
}
      	\subfloat{
%
%
%

\definecolor{mycolor1}{rgb}{0,0.75,0.25}
\definecolor{mycolor2}{rgb}{0.5,1,0}
\definecolor{mycolor3}{rgb}{1,0.5,0.5}
\definecolor{mycolor4}{rgb}{1,0,1}

\begin{tikzpicture}

\begin{axis}[%
width=\figurewidth,
height=\figureheight,
scale only axis,
xmin=0,
xmax=50,
xlabel={SNR [dB]},
ymode=log,
ymin=0.001,
ymax=1000,
yminorticks=true,
ylabel={Function norm-2 error}
]
\addplot [
color=black,
solid,
mark=x,
mark options={solid},
forget plot
]
table[row sep=crcr]{
0 19.1844210262215\\
5 13.3961133585713\\
10 12.1229703663797\\
15 10.3478823236447\\
20 11.4189541424663\\
25 9.74994973642332\\
30 10.3873842428902\\
35 10.3038834259648\\
40 9.74683299013087\\
45 11.2426912726026\\
50 11.6877762535578\\
};
\addplot [
color=blue,
solid,
mark=o,
mark options={solid},
forget plot
]
table[row sep=crcr]{
0 10.1426042339551\\
5 3.26665877261412\\
10 1.89894125104041\\
15 0.959377662711425\\
20 1.04920722317095\\
25 0.571480797264972\\
30 0.82060240260681\\
35 0.99015425246626\\
40 0.683269271027822\\
45 0.941508886422494\\
50 0.80859037480579\\
};
\addplot [
color=mycolor1,
solid,
mark=triangle,
mark options={solid,,rotate=90},
forget plot
]
table[row sep=crcr]{
0 204.46080538696\\
5 30.725936532536\\
10 6.19362794995041\\
15 2.15841863780791\\
20 1.09014783226557\\
25 0.64394629195302\\
30 0.539067353238719\\
35 0.50261807058581\\
40 0.390093209407569\\
45 0.501846688573357\\
50 0.506515301685966\\
};
\addplot [
color=mycolor2,
solid,
mark=star,
mark options={solid},
forget plot
]
table[row sep=crcr]{
0 32.9591492184659\\
5 11.9754661866864\\
10 4.21563965878845\\
15 1.85674651625284\\
20 0.558036507766929\\
25 0.187020571352703\\
30 0.0698749467539456\\
35 0.0422255094872001\\
40 0.0182569220475463\\
45 0.0117726905334601\\
50 0.00927343949202777\\
};
\addplot [
color=purple,
solid,
mark=triangle,
mark options={solid},
forget plot
]
table[row sep=crcr]{
0 31.9326360611928\\
5 11.3977710537475\\
10 4.49101019469745\\
15 1.730447980482\\
20 0.555462698036787\\
25 0.1835368671179\\
30 0.0765899396409224\\
35 0.0355020795697455\\
40 0.0240348344643391\\
45 0.0233501935988081\\
50 0.0137145130506631\\
};
\addplot [
color=mycolor3,
solid,
mark=triangle,
mark options={solid,,rotate=180},
forget plot
]
table[row sep=crcr]{
0 12.964260016709\\
5 5.0885363435249\\
10 2.83403847037637\\
15 1.34121344369979\\
20 1.75346980849462\\
25 1.06402362534858\\
30 0.730725511997513\\
35 1.38033901765886\\
40 0.770532080441904\\
45 1.96981714161408\\
50 1.24720972930223\\
};
\addplot [
color=mycolor4,
solid,
mark=+,
mark options={solid},
forget plot
]
table[row sep=crcr]{
0 127.204714698164\\
5 128.004969178986\\
10 125.395475420815\\
15 130.436903747989\\
20 125.406224252319\\
25 123.809839961227\\
30 126.879480529949\\
35 123.949073818118\\
40 124.855774206824\\
45 128.333778316943\\
50 128.523760619703\\
};
\end{axis}
\end{tikzpicture}%
}\\
      	\subfloat{
%
%
%

\definecolor{mycolor1}{rgb}{0,0.75,0.25}
\definecolor{mycolor2}{rgb}{0.5,1,0}
\definecolor{mycolor3}{rgb}{1,0.5,0.5}
\definecolor{mycolor4}{rgb}{1,0,1}

\begin{tikzpicture}

\begin{axis}[%
width=\figurewidth,
height=\figureheight,
scale only axis,
xmin=0,
xmax=50,
xlabel={SNR [dB]},
ymode=log,
ymin=1e-09,
ymax=10,
yminorticks=true,
ylabel={$\text{\vb-MSE [}\mu{}\text{s}^\text{2}\text{]}$}
]
\addplot [
color=black,
solid,
mark=x,
mark options={solid},
forget plot
]
table[row sep=crcr]{
0 0.0103612751016462\\
5 3.57507004370359e-05\\
10 0.041005139722553\\
15 0.00643850551946805\\
20 3.44539778600639e-05\\
25 3.33613346829362e-05\\
30 0.000473072278603548\\
35 3.38982774094751e-05\\
40 0.0851460495257903\\
45 0.021110896898201\\
50 0.0152706638254656\\
};
\addplot [
color=blue,
solid,
mark=o,
mark options={solid},
forget plot
]
table[row sep=crcr]{
0 3.45961035754669e-06\\
5 1.32138235055132e-06\\
10 4.8798411166857e-07\\
15 2.07546319568733e-07\\
20 1.45073612045866e-07\\
25 1.12591988912534e-07\\
30 1.03039299183924e-07\\
35 1.00915012877699e-07\\
40 1.02922895271029e-07\\
45 1.00183509299143e-07\\
50 9.81165388296545e-08\\
};
\addplot [
color=mycolor1,
solid,
mark=triangle,
mark options={solid,,rotate=90},
forget plot
]
table[row sep=crcr]{
0 2.31224695436979\\
5 0.232605203161995\\
10 9.6024009159866e-06\\
15 6.37535716066527e-06\\
20 6.8414085450848e-06\\
25 6.49175251981335e-06\\
30 6.44128617938526e-06\\
35 5.75688885723374e-06\\
40 7.45134224537558e-06\\
45 5.98558708343454e-06\\
50 5.81645349235797e-06\\
};
\addplot [
color=mycolor2,
solid,
mark=star,
mark options={solid},
forget plot
]
table[row sep=crcr]{
0 1.70551954273554e-05\\
5 8.02586642496928e-06\\
10 2.40096331752246e-06\\
15 7.76191443100289e-07\\
20 2.55856205704007e-07\\
25 9.4895765981357e-08\\
30 2.85586885480333e-08\\
35 2.01847223051574e-08\\
40 1.62670756341509e-08\\
45 1.37024642680654e-08\\
50 1.46740812691267e-08\\
};
\addplot [
color=purple,
solid,
mark=triangle,
mark options={solid},
forget plot
]
table[row sep=crcr]{
0 1.54816582303477e-05\\
5 5.05919365703197e-06\\
10 1.94388945735143e-06\\
15 5.92579657110546e-07\\
20 1.96739786931074e-07\\
25 6.76280993156294e-08\\
30 2.38599559438619e-08\\
35 1.65166441750828e-08\\
40 1.03258104720845e-08\\
45 8.9415820370462e-09\\
50 8.30071703490751e-09\\
};
\addplot [
color=mycolor3,
solid,
mark=triangle,
mark options={solid,,rotate=180},
forget plot
]
table[row sep=crcr]{
0 7.01289762299004e-06\\
5 3.3793779375124e-06\\
10 1.47242565949909e-06\\
15 6.05318567444513e-07\\
20 2.77985015084821e-07\\
25 1.23000953524472e-07\\
30 5.79152698856523e-08\\
35 3.49855169144193e-08\\
40 1.56874987038305e-08\\
45 9.97619616358794e-09\\
50 7.15348841983327e-09\\
};
\addplot [
color=mycolor4,
solid,
mark=+,
mark options={solid},
forget plot
]
table[row sep=crcr]{
0 9.34266630411898\\
5 8.64843572044062\\
10 8.90812753261971\\
15 8.47342653462019\\
20 8.43114163978945\\
25 8.50738464074901\\
30 9.15761963581044\\
35 8.93767310727025\\
40 8.95984160726948\\
45 10.328132608468\\
50 9.51675110752285\\
};
\end{axis}
\end{tikzpicture}%
}
      	\subfloat{
%
%
%

\definecolor{mycolor1}{rgb}{0,0.75,0.25}
\definecolor{mycolor2}{rgb}{0.5,1,0}
\definecolor{mycolor3}{rgb}{1,0.5,0.5}
\definecolor{mycolor4}{rgb}{1,0,1}

\begin{tikzpicture}

\begin{axis}[%
width=\figurewidth,
height=\figureheight,
scale only axis,
xmin=0,
xmax=50,
xlabel={SNR [dB]},
ymode=log,
ymin=1e-09,
ymax=10,
yminorticks=true,
ylabel={$\text{\vb-MSE [}\mu{}\text{s}^\text{2}\text{]}$}
]
\addplot [
color=black,
solid,
mark=x,
mark options={solid},
forget plot
]
table[row sep=crcr]{
0 0.631697866177638\\
5 0.309590319893545\\
10 0.287528017373601\\
15 0.148497795982915\\
20 0.148312462526754\\
25 0.0366867259795568\\
30 0.100887928022646\\
35 0.131743592331803\\
40 0.0893078005451133\\
45 0.10408046696074\\
50 0.135960686036069\\
};
\addplot [
color=blue,
solid,
mark=o,
mark options={solid},
forget plot
]
table[row sep=crcr]{
0 0.455136426092063\\
5 0.15090874592786\\
10 0.000133995316273908\\
15 2.84485749990843e-06\\
20 3.12600055743397e-06\\
25 1.78815704573426e-06\\
30 2.72242339696125e-06\\
35 2.64422992251789e-06\\
40 2.40418776583238e-06\\
45 2.87438059331533e-06\\
50 2.52321563713814e-06\\
};
\addplot [
color=mycolor1,
solid,
mark=triangle,
mark options={solid,,rotate=90},
forget plot
]
table[row sep=crcr]{
0 4.34487776814635\\
5 0.454383235980653\\
10 0.00762061917261624\\
15 7.05930418491443e-06\\
20 6.33974322673672e-06\\
25 6.06783256167559e-06\\
30 6.37604293263457e-06\\
35 7.10908892818089e-06\\
40 6.15274131331773e-06\\
45 6.2809218747564e-06\\
50 5.75174302353097e-06\\
};
\addplot [
color=mycolor2,
solid,
mark=star,
mark options={solid},
forget plot
]
table[row sep=crcr]{
0 0.542269156331818\\
5 0.0695018133346639\\
10 7.66140959138778e-06\\
15 3.92221811048305e-06\\
20 1.86632661259462e-06\\
25 3.25969458136784e-07\\
30 2.13074226132936e-07\\
35 7.78885098431885e-08\\
40 4.64351306528257e-08\\
45 2.29443300804067e-08\\
50 2.13465633887909e-08\\
};
\addplot [
color=purple,
solid,
mark=triangle,
mark options={solid},
forget plot
]
table[row sep=crcr]{
0 0.448378801013336\\
5 0.151065971610487\\
10 0.000133482677356114\\
15 3.05998929339704e-06\\
20 1.92692337073567e-06\\
25 3.23304032293948e-07\\
30 1.79726073871668e-07\\
35 6.71585407614088e-08\\
40 6.56491749687121e-08\\
45 4.76790836014382e-08\\
50 2.38000579236012e-08\\
};
\addplot [
color=mycolor3,
solid,
mark=triangle,
mark options={solid,,rotate=180},
forget plot
]
table[row sep=crcr]{
0 0.476855174387573\\
5 0.190007276667867\\
10 0.000127813732290257\\
15 0.000114456861733394\\
20 4.43580097011683e-06\\
25 0.0158986222363453\\
30 1.89777613084437e-06\\
35 4.16773815492888e-06\\
40 3.55126698095715e-06\\
45 0.0102409546219354\\
50 8.56314860463612e-05\\
};
\addplot [
color=mycolor4,
solid,
mark=+,
mark options={solid},
forget plot
]
table[row sep=crcr]{
0 9.45391871887842\\
5 9.455425997743\\
10 8.40340982864713\\
15 9.10234875513059\\
20 9.04804797303519\\
25 9.12597960538903\\
30 10.1976078789145\\
35 8.42060995557122\\
40 9.2753103797861\\
45 9.24010533286298\\
50 8.9289764099808\\
};
\end{axis}
\end{tikzpicture}%
}
        \caption{Function error (top) and estimator precision (bottom) with overlapping pulses 
        and noise folding for the TDE problem. Left: $\kappa=1$; right: $\kappa=0.4$.}
        \label{fig:tde_K3_overlapping_nf_pulses}
        \vspace{-5mm}
\end{figure*}

In our final numerical experiment we investigate the effect of signal noise in
the received signal, rather than measurement noise. Signal noise introduces
noise folding \cite{Castro2011,Treichler2011}, which decreases reconstruction
performance. Signal noise occurs when the signal models is: $\vy =
\mA(\mD\vx + \vn) + \vw$, where $\vn$ is the signal noise and $\vw$ is the
measurement noise. In the experiments so far we have only considered measurement
noise, but now we focus on the signal noise and set the measurement noise to
zero. The estimator performance for $\kappa=1$, i.e., no subsampling, and
$\kappa=0.4$ is shown in \figref{fig:tde_K3_overlapping_nf_pulses}.
As can be seen in the top two figures all the estimators' signal reconstruction
are affected by noise folding when CS is used. However, in the bottom two
figures we see that for the parameter estimation some of the algorithms are
still able to estimate the translation parameter at a similar precision as
without subsampling. The greedy algorithms are most heavily affected by the
noise folding, whereas the convex optimization based algorithms are less
affected. As shown in \cite{Treichler2011} noise folding may be remedied by
using quantization. In that work, the authors postulate that when a receiver
uses, e.g., half as many samples as a classical receiver, it may instead use
twice as many bits for quantization. This is also demonstrated in
\cite{Fyhn2013b} for a spread spectrum receiver.

\section{Discussion}
\label{sec:discussion}
With our numerical experiments, we show that the proposed CCBP and, with high
enough measurement rates, the PaIBOMP+CCBP algorithm outperform 
existing TDE algorithms in terms of estimation precision. If the pulses are known to 
be well separated, the PoIBOMP algorithm attains the best estimation precision, 
while having very low computational complexity. If the pulses cannot be assumed to 
be well separated, it is better to use the pure convex optimization algorithm CCBP or 
the hybrid PaIBOMP+CCBP to attain the best estimation precision. 

Our experiment in Section V.A studied the effect on the estimation performance of the parameter spacing $\Delta$ of the dictionary.
While reducing the value of $\Delta$ can potentially improve the parameter estimation resolution, such  reduction results in an increase of the coherence in the dictionary.
We have experimentally verified that the higher coherence present in the FE problem in comparison with the TDE problem impacts the estimation performance; additionally, we verified this higher coherence by evaluating the spark of the two dictionaries used.
The spark formulation also allowed us to clearly show the reason for the CBP formulation being successful for positive-valued coefficients but unsuccessful for arbitrary complex coefficients; recall that the spark is a more stringent measure of coherence in terms of linear dependence of the vectors. A similar analysis, which determines the necessary trade-off point between a unique sparse solution for the convex optimization algorithm and sufficiently small approximation error for the circle approximation on the manifold, is necessary if the algorithms presented here are to be applied to additional problems.

At lower subsampling ratios the algorithms that achieve the best performance are CCBP or
TDE-MUSIC. In our experiments CCBP attained the best estimation precision;
however, it is also significantly more computationally complex. It may be
possible to reduce this complexity through a more judicious formulation of
solvers for the proposed optimization. The proposed modified optimization
problem introduces many new variables to be able to capture the full signal
information, but it may be possible to decrease this number with a smarter
problem formulation.

In the last numerical experiment, we investigated the estimators' performance
when the observations feature signal noise instead of measurement noise. This
results in noise folding which has been shown before to severely affect signal
reconstruction. In our experiments we see that the greedy algorithms are highly
sensitive to such noise folding, while TDE-MUSIC, CCBP and PaIBOMP+CCBP are less
sensitive. This is an interesting, novel result as it indicates that compressive parameter estimation may be less affected by noise folding than compressive signal reconstruction.

Finally, we point out that while the use of accurate interpolation models can alleviate 
the issues due to parameter sampling when parametric dictionaries are used, there 
are other inherent limitations to sparsity-leveraging approaches that our approaches 
are subject to as well. This is illustrated by the FE problem, in which the 
inclusion of additional elements in the parametric dictionary to expand to 
complex amplitudes severely affects the conditioning of the resulting CS matrix.

\section{Conclusion}
\label{sec:conclusion}
Our work in this paper shows that compressive sensing for the class of sparse
translation-invariant signals allows for a lower sampling rate and that the use
of polar interpolation increases the estimation precision. The cost in terms of
computational complexity is a trade-off in terms of the desired estimation
precision and whether it is known if the signal pulses are well-separated or
not.

The results presented here are intended as a general exploration of three parameter values: estimator precision, sampling rate and computational complexity in the range of their trade-off.
Before the presented algorithms may be employed for a given scenario this therefore requires an analysis to find the best parameter values $\zeta$ and $\lambda$ and the best choice of algorithm.

Future work will investigate how the algorithms proposed here may be used in other scenarios and also if they may be extended to multiple parameter estimation problems, such as in radar and GPS localization where both a frequency and a delay must be estimated~\cite{Herman09,Misra13}.


\end{document}